\documentclass[iop,numberedappendix,twocolumn]{emulateapj}
\usepackage{amsmath, amsfonts, amssymb}

\shorttitle{GRB~140430A}
\shortauthors{Kopa\v c et al.}

\begin{document}
\title{Limits on optical polarization during the prompt phase of GRB~140430A}
\author{D.~Kopa\v c\altaffilmark{1,2}, C.~G.~Mundell\altaffilmark{2,3}, J.~Japelj\altaffilmark{1}, D.~M.~Arnold\altaffilmark{2}, I.~A.~Steele\altaffilmark{2}, C.~Guidorzi\altaffilmark{4}, S.~Dichiara\altaffilmark{4}, S.~Kobayashi\altaffilmark{2}, A.~Gomboc\altaffilmark{1}, R.~M.~Harrison\altaffilmark{5}, G.~P.~Lamb\altaffilmark{2}, A.~Melandri\altaffilmark{6}, R.~J.~Smith\altaffilmark{2}, F.~J.~Virgili\altaffilmark{2}, A.~J.~Castro-Tirado\altaffilmark{7,8}, J.~Gorosabel\altaffilmark{7,9}\textdagger, A.~J{\" a}rvinen\altaffilmark{10}, R.~S{\' a}nchez-Ram{\' i}rez\altaffilmark{7}, S.~R.~Oates\altaffilmark{7}, M.~Jel{\' i}nek\altaffilmark{11}}
\altaffiltext{1}{Department of Physics, Faculty of Mathematics and Physics, University of Ljubljana, Jadranska 19, 1000 Ljubljana, Slovenia.}
\altaffiltext{2}{Astrophysics Research Institute, Liverpool John Moores University, 146 Brownlow Hill, Liverpool, L3 5RF, UK.}
\altaffiltext{3}{Department of Physics, University of Bath, Claverton Down, Bath, BA2 7AY, UK.}
\altaffiltext{4}{Department of Physics and Earth Sciences, University of Ferrara, via Saragat 1, 44122, Ferrara, Italy.}
\altaffiltext{5}{Department of Astrophysics, School of Physics and Astronomy, Tel Aviv University, 69978 Tel Aviv, Israel.}
\altaffiltext{6}{INAF -- Osservatorio Astronomico di Brera, via E. Bianchi 46, 23807 Merate (LC), Italy.}
\altaffiltext{7}{Instituto de Astrof\'{i}sica de Andaluc\'{i}a (IAA-CSIC), Glorieta de la Astronomia s/n, E-18008 Granada, Spain.}
\altaffiltext{8}{Departamento de Ingeniar\'{i}a de Sistemas y Autom\'{a}tica, E.T.S.I. Industriales, Universidad de M\'{a}laga, C/. Doctor Ortiz Ramos s/n, Campus de Teatinos, E-29071 M\'{a}laga, Spain.}
\altaffiltext{9}{Grupo de Ciencias Planetarias, Escuela Superior de Ingenieros, F\'{i}sica Aplicada I, Alameda de Urquijo s/n, E-48013 Bilbao, Spain.}
\altaffiltext{10}{AIP -- Leibniz-Institut f\"{u}r Astrophysik Potsdam, An der Sternwarte 16, 14482 Potsdam, Germany.}
\altaffiltext{11}{ASU-CAS -- Astronomical Institute of the Czech Academy of Sciences, Fri\v cova 298, 251 65 Ond\v{r}ejov, Czech Republic.}
\altaffiltext{\textdagger}{Deceased April 2015}
\email{drejc.kopac@fmf.uni-lj.si}

\begin{abstract}

\par Gamma-ray burst GRB~140430A was detected by the \textit{Swift} satellite and observed promptly with the imaging polarimeter RINGO3 mounted on the Liverpool Telescope, with observations beginning while the prompt $\gamma$-ray emission was still ongoing. In this paper, we present densely sampled (10-second temporal resolution) early optical light curves in 3 optical bands and limits to the degree of optical polarization. We compare optical, X-ray and gamma-ray properties and present an analysis of the optical emission during a period of high-energy flaring. The complex optical light curve cannot be explained merely with a combination of forward and reverse shock emission from a standard external shock, implying additional contribution of emission from internal shock dissipation. We estimate an upper limit for time averaged optical polarization during the prompt phase to be as low as $P < 12\% \,(1\sigma)$. This suggests that the optical flares and early afterglow emission in this GRB are not highly polarized. Alternatively, time averaging could mask the presence of otherwise polarized components of distinct origin at different polarization position angles. 

\end{abstract}


\maketitle

\section{Introduction}

\par Gamma-ray bursts (GRBs) are powerful cosmic explosions, first identified by the detection of short flashes of gamma-ray emission by military satellites in the 1960s \citep{klebesadel1973} and, today, thought to represent the end product of massive stellar core-collapse or the merger of compact objects (e.g., \citealt{meszaros2006, vedrenne2009, gomboc2012}). In addition to being interesting in their own right -- as black hole:jet systems with ultra-relativistic expansion speeds and potentially strong magnetic fields -- GRBs are also among the most distant known objects in the Universe and thus act as probes of the early Universe \citep{tanvir2009, salvaterra2009, cucchiara2011}.

\par In the standard fireball model of GRBs (e.g., \citealt{piran1999}), prompt gamma-ray emission is produced by internal shocks \citep{reesmeszaros1994}, and the long-lasting afterglow emission at longer wavelengths is produced by external shocks when relativistic ejecta collide and are decelerated by the surrounding circumburst material \citep{reesmeszaros1992, meszaros1997}. Despite the overall success of this framework, the prompt emission mechanism is still poorly known, and internal shocks remain an inefficient mechanism for the conversion of kinetic to radiated energy (e.g., \citealt{beloborodov2005, reesmeszaros2005, zhangyan2011, axelsson2015, beniamini2015}). In addition, although the generally accepted external shock model works well for smoothly fading late time ($\sim$ days post burst) afterglows, observations of early afterglow light curves in the first minutes to hours after the burst, especially in the era of the \textit{Swift} satellite \citep{gehrels2004}, show an unexpected wealth of variety, attributed to a range of mechanisms including internal and external shocks, long-lived central engines and double jet structures (e.g., \citealt{monfardini2006, mundell2007b, gomboc2008, melandri2008, melandri2010, virgili2013, kopac2013, japelj2014, depasquale2015}).

\par Bursts with longer-lasting prompt emission or those with very bright optical afterglows -- detectable by small telescopes with wide-fields of view -- have provided the best chance to detect longer wavelength emission during the prompt phase. Although the sample is still relatively small, an increasing number of flares at wavelengths below the gamma-ray band have been detected for a number of GRBs. Similar observations have also been obtained for X-ray flashes\footnote{For discussion on X-ray flares see e.g., \citealt{burrows2005, obrien2006, chincarini2007, margutti2010}.}$^\mathrm{,}$\footnote{For optical flares/rebrightenings see e.g., \citealt{akerlof1999, blake2005, vestrand2005, page2007, mundell2007a, racusin2008, thoene2010, guidorzi2011, gendre2012, gendre2013, kopac2013, virgili2013, elliott2014, greiner2014, vestrand2014, nappo2014}.}$^\mathrm{,}$\footnote{For X-ray flashes see e.g., \citealt{boer2006, kruehler2009, guidorzi2009}.}.

\par The origin and connection between different observed spectral components in the prompt phase remains problematic. Prompt GRB light curves often show rapid variability, which can be explained within internal shock dissipation model \citep{kobayashi1997}. Alternative scenarios have been suggested such as inverse-Compton scattering (e.g. \citealt{panaitescu2008}), large-angle emission (e.g. \citealt{kumar2008}), structured outflow (e.g. \citealt{panaitescuvestrand2008}), anisotropic emission (e.g. \citealt{beloborodov2011}), magnetic reconnection (e.g. \citealt{zhangyan2011}), etc. Particularly, the study of early time optical emission and its polarization properties has provided valuable insight into GRB emission mechanisms (e.g., \citealt{yost2007, yost2007b, kopac2013}), jet composition and magnetic field properties (e.g., \citealt{mundell2007a, steele2009, gomboc2009, mundell2013, japelj2014, king2014}), but the major complications remain observational limitations. Specifically, the prompt optical emission is in most cases relatively faint and, due to this, temporal resolution in the optical band is often inadequate for direct comparison with gamma-ray light curves, resulting in losing information on intrinsic complexity which imprints the central engine behavior. In short, observing and understanding the prompt and early afterglow emission of GRBs remains technically and theoretically challenging. Therefore, GRBs for which multi-wavelength data is obtained simultaneously with the prompt gamma-ray emission and at high temporal cadence are of particular value.

\par Here we present exquisitely sampled 3-band simultaneous multicolor light curves of long-duration GRB~140430A, observed with the RINGO3 polarimeter \citep{arnold2012} on the 2-m robotic Liverpool Telescope (LT; \citealt{steele2004, guidorzi2006}). The prompt gamma-ray emission lasted for 200~seconds, with fainter emission detected as late as 575~seconds after the burst. The optical observations with LT began just 124~seconds after the onset of the burst and were contemporaneous with the gamma and X-ray flares at this time. We present the observations and data reduction in Section \ref{sect:obsred}, analysis of the light curves in Section \ref{sect:analysis}, discussion of our results in the context of the fireball model in Section \ref{sect:discussion} and conclude in Section \ref{sect:conclusions}. Throughout the paper the convention $F_\nu (t) \propto t^{-\alpha}\,\nu^{-\beta}$ is used to describe the flux density. $\Lambda$CDM cosmology is assumed with parameters $H_0 = 67.3\,\mathrm{km\,s^{-1}\,Mpc^{-1}}$, $\Omega _\Lambda = 0.68$, $\Omega _\mathrm{M} = 0.32$ \citep{planck2014}. Best fit parameters are given at $1\sigma$ confidence level, except when stated otherwise. Times are given with respect to GRB trigger time $t_0$.

\section{Observations and reduction}
\label{sect:obsred}

\subsection{Swift}

\par On 2014 April 30, at $t_0 = \mathrm{20:33:36 \, UT}$, the Burst Alert Telescope (BAT; \citealt{barthelmy2005}) onboard the \textit{Swift} satellite triggered on the long GRB~140430A and immediately slewed to the burst \citep{siegel2014}. The BAT gamma-ray light curve shows a multi-peaked structure with two intense peaks: the first one starts at $\sim -10\,\mathrm{s}$ and ends at $\sim 10\,\mathrm{s}$, and the second softer one starts at $\sim 140\,\mathrm{s}$ and ends at $\sim 200\, \mathrm{s}$ post trigger. There are at least two slightly fainter and softer peaks centered at $\sim 25\,\mathrm{s}$ and $\sim 575\,\mathrm{s}$. The $\mathrm{T_{90}}$ ($15-350\,\mathrm{keV}$) of $174 \pm 4\,\mathrm{s}$, fluence ($15-150\,\mathrm{keV}$) of $1.1 \pm 0.2 \times 10^{-6}\,\mathrm{erg/cm^2}$ and time-averaged spectrum power-law index $\Gamma$ of $2.0 \pm 0.2$ \citep{krimm2014} put this GRB towards the long--soft end of \textit{Swift} GRBs \citep{sakamoto2011}. Fitting the BAT time-averaged spectrum with a typical Band function assuming $\alpha = -1$ and $\beta = -2.3$ gives $E_\mathrm{peak}^\mathrm{obs} \sim 20\,\mathrm{keV}$ and, moving to $1 - 10^4\,\mathrm{keV}$ in the host rest frame at redshift of $z=1.6$, gives isotropic-equivalent energy of $E _ {\gamma , \mathrm{iso}} = (1.3 \pm 0.4) \times 10^{52} \,\mathrm{erg}$.

\par The \textit{Swift} X-ray Telescope (XRT; \citealt{burrows2005a}) began follow-up observations at $50.8\,\mathrm{s}$, while gamma-ray emission was still ongoing. A bright and uncatalogued fading X-ray source has been detected at RA(J2000) = 06$^\mathrm{h}$51$^\mathrm{m}$44{\hbox{$.\!\!^{\rm s}$}6, Dec.(J2000) = +23$^\circ$01$^\prime$25$^{\prime \prime}$ with an enhanced $90\%$ confidence uncertainty of $1.9^{\prime \prime}$ \citep{evans2014}. The X-ray light curve at early times is dominated by at least 3 bright flares, centered at $154 \,\mathrm{s}$, $171 \,\mathrm{s}$ and $222 \,\mathrm{s}$, as seen on Figure \ref{fig:xrtlc}. The first two flares track the gamma-ray light curve both temporally and in brightness (see Figures \ref{fig:xrtlc} and \ref{fig:lc}). The last flare is followed by a steep decay. From $\sim 500\,\mathrm{s}$ to $\sim 3900\,\mathrm{s}$ a data gap in X-ray is due to the Earth limb constraint. At later times the X-ray light curve shows a decay until $\sim 10^5\,\mathrm{s}$.

\begin{figure}[!h]
\begin{center}
\includegraphics[width=1\linewidth]{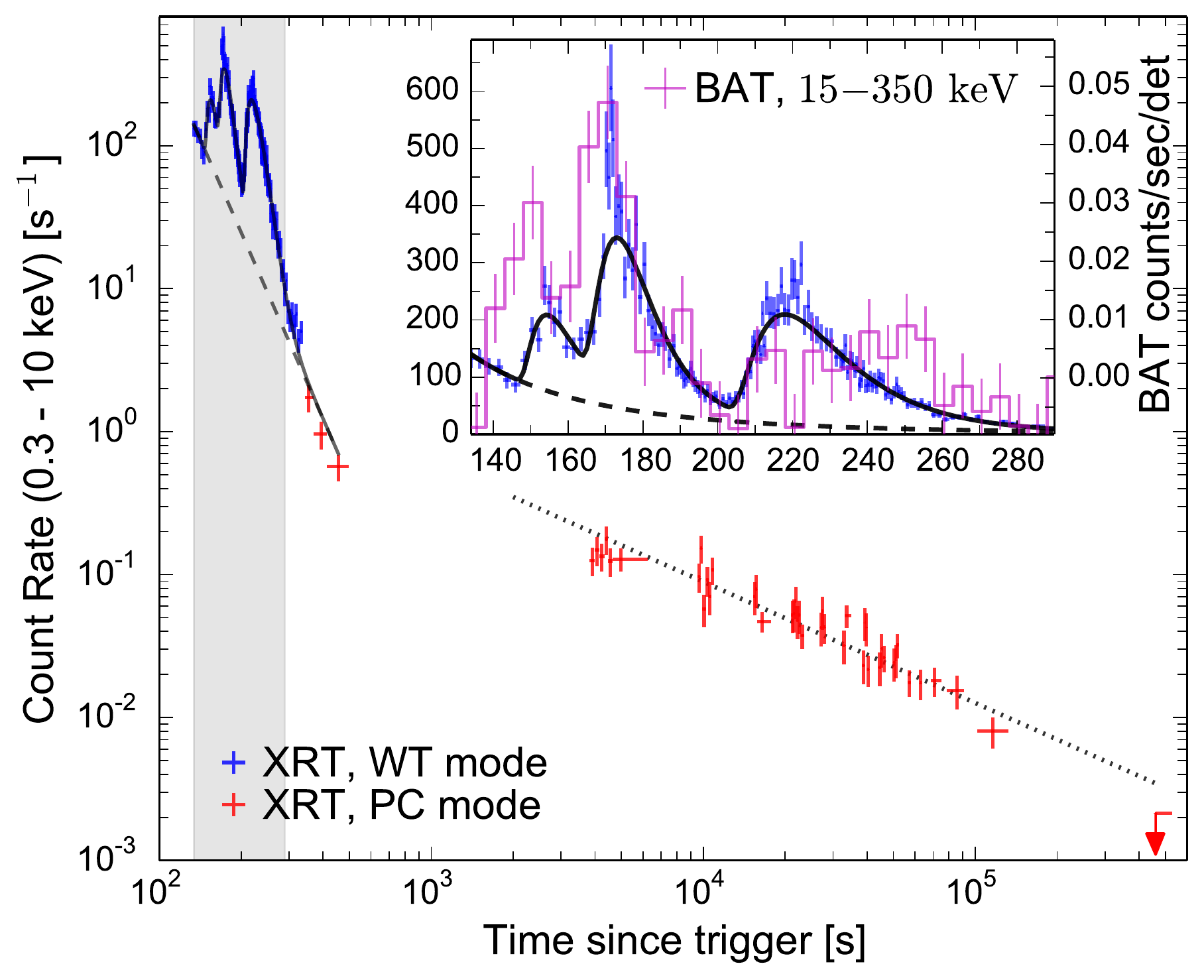} 
\caption{\label{fig:xrtlc} GRB~140430A X-ray light curve obtained from the XRT repository \citep{evans2009}. Inset plot shows early time behavior in linear scale, together with the gamma-ray light curve (violet histogram, $5\,\mathrm{s}$ uniform binning) obtained from the BAT instrument. Black solid and dash lines represent the fitted model as described in Section \ref{sect:xray}.}
\end{center}
\end{figure}

\par When transforming the X-ray light curve obtained from the XRT light curve repository \citep{evans2009} from flux to flux density units, we took into account strong spectral evolution at early times and by assuming a power-law spectrum with a given spectral index for each point (see Figure \ref{fig:lc_beta}), we calculated the flux density at $10\,\mathrm{keV}$ by integrating energy spectrum in the $0.3 - 10\,\mathrm{keV}$ interval. At later times, we instead assumed an average spectral index for every point (see Section \ref{sect:xray}).

\par The \textit{Swift} Ultraviolet/Optical Telescope (UVOT; \citealt{roming2005}) began observing the GRB field at $183\,\mathrm{s}$ and found a candidate afterglow at the position consistent with the XRT, with estimated magnitude in the UVOT White filter of $18.17 \pm 0.09\,\mathrm{mag}$ \citep{breeveld2014}. 

\subsection{Liverpool Telescope}

\par LT responded to the \textit{Swift} GRB alert automatically and started observations with RINGO3 instrument at $\mathrm{20:35:40 \, UT}$, i.e., $124\,\mathrm{s}$ after the BAT trigger \citep{melandri2014}. Fast response to the trigger resulted in obtaining the optical observations contemporaneously with the ongoing gamma-ray and X-ray emission during high-energy flares at early times (Figure \ref{fig:lc}). Observations continued for the next hour, during which the majority of the observations were performed with RINGO3, with a $6\times 10\,\mathrm{s}$ sequence of IO:O observations with an SDSS r' filter inserted at $\sim 33\,\mathrm{min}$ to allow for real-time afterglow identification during the observations.

\par RINGO3 is a novel 3-band fast-readout optical imaging polarimeter, which uses a polaroid that rotates once per second. By analyzing relative intensities at 8 different orientations of the polaroid, the polarization for each source in the image can be measured, while summing the data from all rotation angles allows derivation of the total flux of each source. Short $125\,\mathrm{ms}$ exposures and zero read-out noise allows optimization of frame co-adding in the data post-processing stages. In addition, a light entering the instrument and passing the rotating polaroid is split into 3 beams using a pair of dichroic mirrors, and simultaneously imaged using 3 separate EMCCD cameras. The wavelength bands\footnote{http://telescope.livjm.ac.uk/TelInst/Inst/RINGO3/} are determined by dichroics: V ranges from $\sim 350 - 640 \,\mathrm{nm}$, R ranges from $\sim 650 - 760\,\mathrm{nm}$ and I ranges from $\sim 770 - 1000\,\mathrm{nm}$. The transmittance and response in UV and IR part of these bands is further affected by the camera quantum efficiency. On average, the wavelength bands are approximately equivalent to VRI Johnson-Cousins photometric system. The wavelength range covered is largest for the V-equivalent band, thereby providing the highest signal-to-noise ratio of the 3 cameras.

\par Data from RINGO3 were automatically stacked to produce frames with $10\,\mathrm{s}$ and $60\,\mathrm{s}$ exposures. Frames were cleaned using the Singular Spectrum Analysis decomposition procedures \citep{golyandina2013}, to remove vignetting and fringing. All together, the LT dataset consisted originally of $65$ frames in each of the 3 color bands. To enhance signal to noise ratio at later times when the afterglow brightness was below $\sim 18\,\mathrm{mag}$, frames with $60\,\mathrm{s}$ exposure were co-added from $\sim 10^3\,\mathrm{s}$ onwards. 

\par Because no standard stars' fields were available for the night of the GRB observations, photometric calibration was performed relative to the magnitudes of $5$ non-saturated USNO-B1.0 stars in the field, using standard aperture photometry procedures. USNO I cataloged magnitudes were used to calibrate I-equivalent RINGO3 band, R2 magnitudes for R-equivalent RINGO3 band, and approximated $\mathrm{V} \sim 0.444 \times \mathrm{B1} + 0.556\times \mathrm{R2}$ magnitudes for V-equivalent RINGO3 band\footnote{The V band estimate using USNO B and R magnitudes is very crude; see www.aerith.net/astro/color\_conversion/JG/USNO-B1.0.html (Greaves 2003).}. While USNO-B1.0 magnitudes provide relatively poor absolute photometric accuracy ($\pm 0.3 \,\mathrm{mag}$), the photometric stability throughout the observations was very good as shown the variability in zero points being $< 0.1\,\mathrm{mag}$ in each RINGO3 band. 

\subsection{BOOTES, IAC, OSN, STELLA, GTC}

The afterglow of GRB~140430A was observed by the following facilities: 

\par The $0.6\,\mathrm{m}$ TELMA robotic telescope at the BOOTES-2 astronomical station at IHSM/UM-CSIC La Mayora, Spain \citep{castrotirado2012} started observing the afterglow with the COLORES imaging spectrograph at $30\,\mathrm{s}$ after the BAT trigger with r$^{\prime}$- and i$^{\prime}$- band filters. The detected peak magnitude was r$^{\prime}$ $\sim$ 16.1 mag. Due to poor observing conditions, the observations resulted in poor image quality. Nevertheless we could extract the r$^{\prime}$ -band magnitude for two early epochs, which are represented in Figure \ref{fig:earlyBAT_BOOTES}. The first epoch is marginally contemporaneous with the second gamma-ray peak in BAT light curve. Due to significantly poorer temporal sampling of BOOTES data compared to the later optical data, we did not include the BOOTES points in the subsequent detailed light curve analysis.

\begin{figure}[!h]
\begin{center}
\includegraphics[width=1\linewidth]{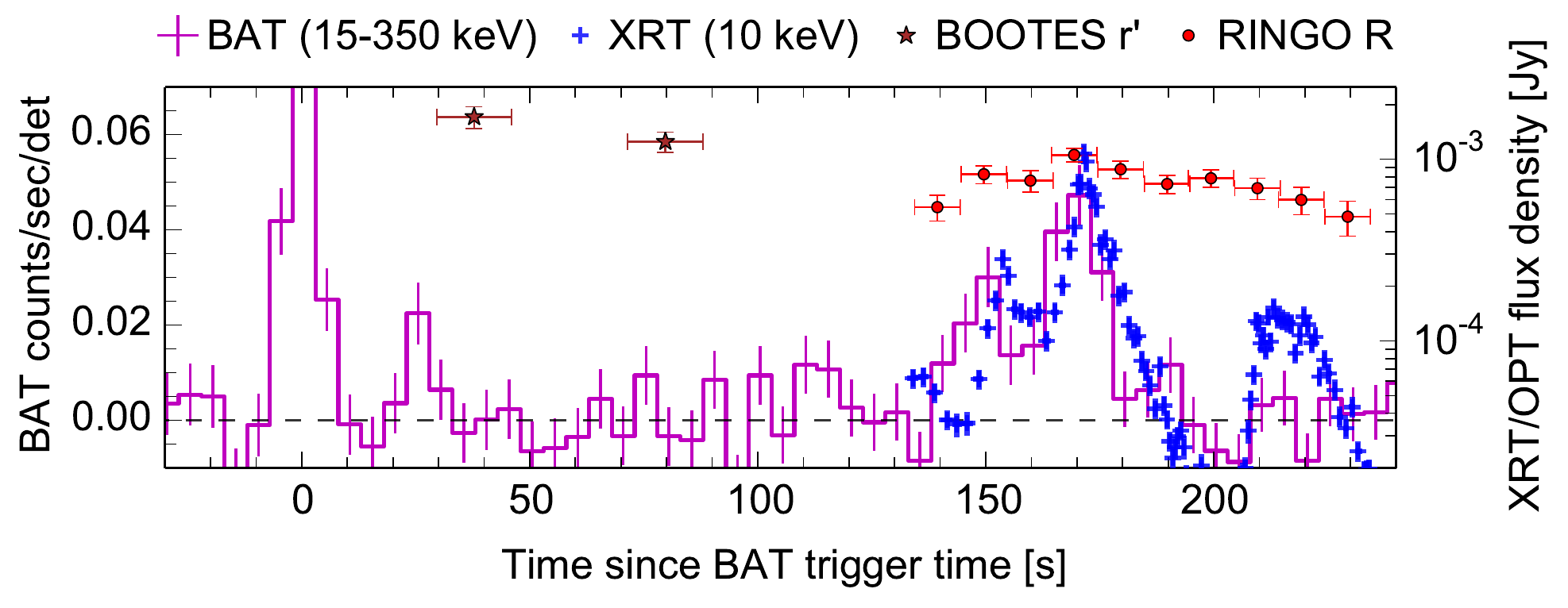} 
\caption{\label{fig:earlyBAT_BOOTES} GRB~140430A early light curve in gamma-ray (BAT), X-ray (XRT), and optical bands, including 2 BOOTES epochs.}
\end{center}
\end{figure}

\par IAC-80 $0.82\,\mathrm{m}$ telescope (Observatorio del Teide) observed the afterglow between $25.0\,\mathrm{min}$ and $40.9\,\mathrm{min}$, providing $3\times 300\,\mathrm{s}$ sequence of observations in BVR-band filters \citep{gorosabel2014}. 

\par OSN $1.5\,\mathrm{m}$ telescope (Observatorio de Sierra Nevada) observed the afterglow between $6.7\,\mathrm{min}$ and $78\,\mathrm{min}$, providing all together $280$ frames in I band, with gradually extending the exposure time from $5\,\mathrm{s}$ to $60\,\mathrm{s}$. To enhance signal to noise at early times we coadded the frames to similar binning as the LT frames, while at late times we used $240\,\mathrm{s}$ binning. 

\par STELLA-I $1.2\,\mathrm{m}$ robotic telescope (Observatorio del Teide) observed the afterglow between $2.9 \,\mathrm{min}$ and $32.5\,\mathrm{min}$ in r' band filter, providing $52$ frames with $20\,\mathrm{s}$ exposure times. To enhance signal to noise we coadded the frames to $60\,\mathrm{s}$ binning at early times and to $120\,\mathrm{s}$ binning at late times. 

\par The 10.4m Gran Telescopio Canarias (GTC) telescope observed the afterglow at $\sim$3.5 $\times$ 10$^{3}$ s, obtaining an optical spectra with the OSIRIS imaging spectrograph \citep{cepa2000}, using R1000R, R2000B and R2500R grisms. Data was reduced and calibrated on the usual way using IRAF and custom python tools. We clearly detect several absorption lines interpreted as CIV$\lambda$1548 \& $\lambda$1551, AlII$\lambda$1671, FeII$\lambda$2600, MgII$\lambda$2796, MgII$\lambda$2803 and MgI$\lambda$2853 at a common redshift of $z = 1.600 \pm 0.001$, consistent with \citet{kruehler2014}.

\par Frames from IAC, OSN and STELLA telescopes have been calibrated against the same USNO-B1.0 stars and in the same manner as the LT frames. The resulting light curves (LCs) are plotted in Figure \ref{fig:lc}. Comparison between data points from RINGO3 instrument (using dichroics) and data points from other telescopes (using standard filters) confirms that RINGO3 bands are well approximated by VRI-equivalent photometric system. All calibrated magnitudes were later corrected for the Galactic extinction of $\mathrm{E_{B-V}} = 0.12 \,\mathrm{mag}$ in the direction of the burst \citep{schlafly2011}, and converted to flux densities using \citet{fukugita1996}. The complete photometry is available in Table 4.

\begin{figure*}[!ht]
\begin{center}
\includegraphics[width=1\linewidth]{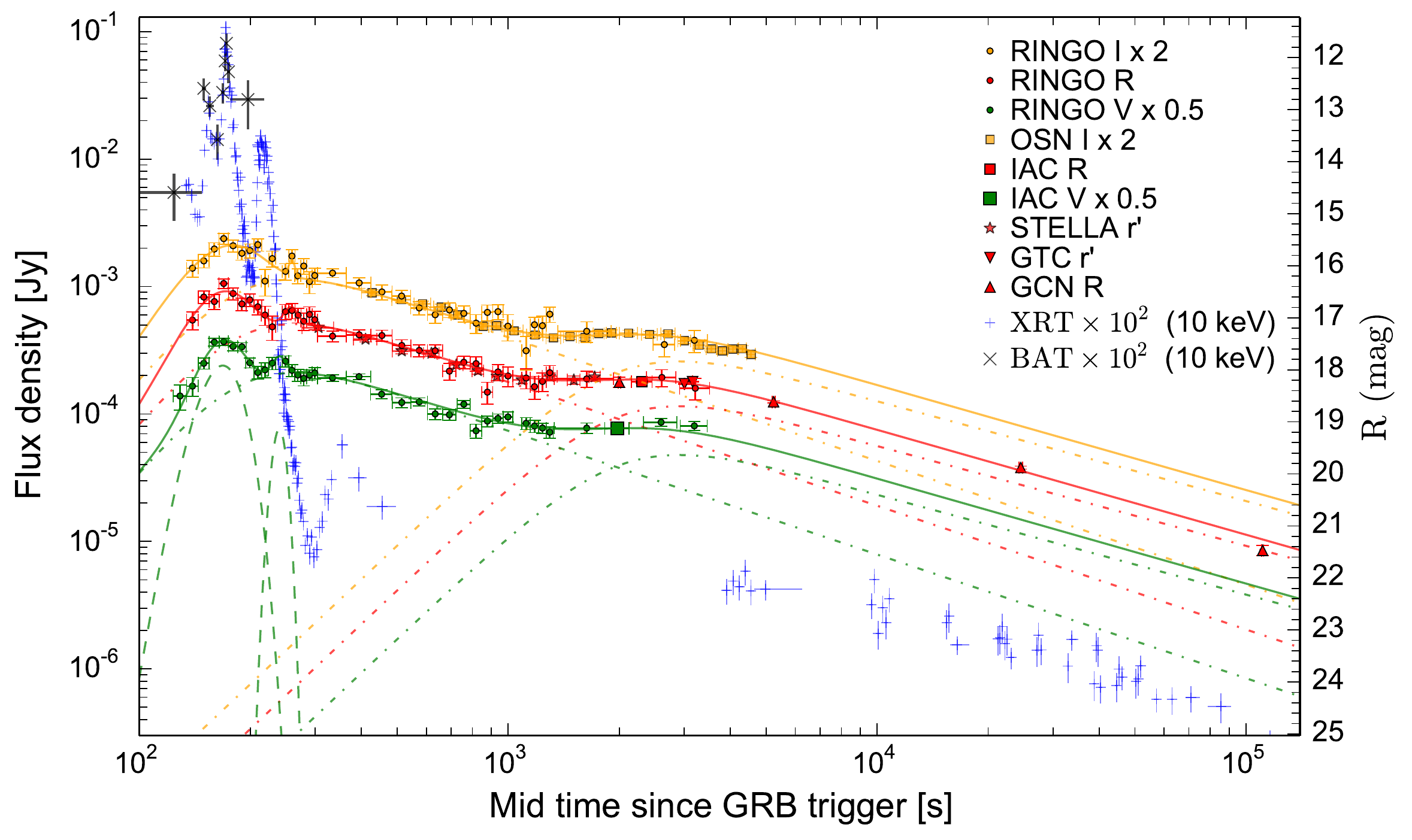} 
\caption{\label{fig:lc} GRB~140430A multi-wavelength light curves in gamma-ray, X-ray and optical bands. Optical data are best modeled (solid lines) using the sum of 2 Beuermann (dashed-dotted lines) profiles to describe broad underlying components and 2 Gaussian (dashed lines) profiles to describe early time flares (Section \ref{sect:optana}).}
\end{center}
\end{figure*}

\section{Analysis}
\label{sect:analysis}

\subsection{X-ray data}
\label{sect:xray}

\par The X-ray light curve shows strong variability and rapid flaring activity at early time. Flaring of the soft gamma-ray emission is also seen at this time, up to $\sim 200\,\mathrm{s}$ post burst, simultaneous with the X-ray flares and showing similar temporal structure in the light curve (Figures \ref{fig:xrtlc} and \ref{fig:lc}). At the end of the steep X-ray decay phase at $\sim 500\,\mathrm{s}$, the X-ray data gap occurs due to Earth occultation and lasts until $4000\,\mathrm{s}$. During this time, the X-ray light curve indicates the likely presence of the canonical X-ray afterglow plateau phase \citep{nousek2006}. From $4000\,\mathrm{s}$ to $\sim 1.3\times 10^5\,\mathrm{s}$ the decay is best described by a broken power-law with poorly constraint break time $t_\mathrm{b} \sim 9700\,\mathrm{s}$, and with decay indices $\alpha _1 = 0.51 \pm 0.32$ pre-break and $\alpha _2 = 0.85 \pm 0.06$ post-break ($\chi ^2 _\mathrm{red} = 0.99$). The X-ray upper limit obtained at $4.6 \times 10^5\,\mathrm{s}$ is not consistent with a simple extrapolation of the late time decay, indicating towards the possible occurrence of a jet break after $10^5 \,\mathrm{s}$. 

\par The time-averaged spectrum formed from the early time data (Windowed Timing -- WT mode, see Figure \ref{fig:xrtlc}) can be well fitted with the absorbed power-law. Fixing the Galactic absorption column to $N_\mathrm{HI}^\mathrm{Gal.} = 2.13\times 10^{21} \, \mathrm{cm^{-2}}$ \citep{willingale2013}, the resulting intrinsic absorption at $z=1.6$ is $N_\mathrm{HI} ^\mathrm{Host}= (3.4 \pm 1.4) \times 10^{21} \, \mathrm{cm^{-2}}$, and the power-law photon index of the spectrum is $\Gamma = \beta + 1 = 2.10 \pm 0.04$ \citep{evans2009}. All values are consistent with the late time data (Photon Counting -- PC mode, see Figure \ref{fig:xrtlc}).

\par To discuss the early time X-ray emission properties, we modeled the early X-ray light curve with a combination of an underlying power-law decay and 3 superimposed bumps described by the Norris profile \citep{norris2005}:
\begin{align}
\label{eq:norris}
F_\mathrm{X} (t) &= F_0 \left( \frac{t}{t_0}\right)^{-\alpha_\mathrm{decay}} \notag \\
& + \sum _{i=1}^{3} F_i \exp{\left(2\sqrt{\frac{\tau _{1,i}}{\tau _{2,i}}}\right)}\,\exp{\left(-\frac{\tau _{1,i}}{\tilde{t}} - \frac{\tilde{t}}{\tau_{2,i}}\right)}\,,
\end{align}
where $F_0$ is the power-law normalization factor and $\alpha_\mathrm{dec}$ is the overall power-law decay index, $F_i$ are normalization constants of superimposed bumps, $\tau _{1,i}$ and $\tau _{2,i}$ are factors determining the shape of each bump, and $\tilde{t}$ is the time measured from the start of each bump as determined from light curve. From parameters of Norris profiles for each bump it is possible to obtain peak times as $t_{\mathrm{peak},i} = \sqrt{\tau_{1,i} \tau_{2,i}}$ and peak durations as $\Delta t _i = \tau _{2,i} \, \sqrt{1 + 4\sqrt{\tau _{1,i} /\tau _{2,i}}}$ \citep{norris2005}. The results of fitting this model (Equation \ref{eq:norris}) to early X-ray light curve are presented in Table \ref{tab:xrayfit}. We note that the second X-ray bump is poorly fitted with the Norris profile as the height of the peak is significantly above the fitted function (see Figure \ref{fig:xrtlc}), but the obtained peak time and width are nevertheless reasonable.

\begin{deluxetable}{cccccc}[!h]
\tablecaption{Early X-ray LC best fit (power-law + 3 Norris peaks) parameters}
\tabletypesize{\footnotesize}
\tablewidth{\linewidth}
\tablehead{
\colhead{Peak} &
\colhead{Interval} &
\colhead{Peak time} & 
\colhead{Duration} &
\colhead{$\Delta t/t$} &
\colhead{$\Delta F/F$} \\
\colhead{} &
\colhead{$[\mathrm{s}]$} &
\colhead{$t_\mathrm{peak}\,[\mathrm{s}]$} & 
\colhead{$\Delta t\,[\mathrm{s}]$} &
\colhead{} &
\colhead{} 
}
\startdata
1 & $144$ - $160$ & $154.6 \pm 1.7$ & $18.7 \pm 4.9$ & $0.12$ & $1.77$ \\
2 & $160$ - $200$ & $173.2 \pm 0.8$ & $22.0 \pm 1.1$ & $0.13$ & $6.47 \,^\ast$ \\
3 & $200$ - $496$ & $218.5 \pm 0.7$ & $37.3 \pm 1.0$ & $0.17$ & $11.42$
\enddata
\tablecomments{Peak times, durations, and flux ratios of 3 X-ray flares obtained from early time fit (see Section \ref{sect:xray}). The underlying power-law component has a decay index of $\alpha_\mathrm{decay} = 4.3 \pm 0.1$. The $\chi ^2 _\mathrm{red}$ of the fit is $\chi ^2 _\mathrm{red} = 1.19$ with $146$ d.o.f. $^\ast$Because the fit underestimates the flux of the second peak (see the text), $\Delta F/F$ for the second peak could be larger by a factor of $\sim 2$.}
\label{tab:xrayfit}
\end{deluxetable}

\par Short variability timescales ($\Delta t/t$) and large amplitude variability ($\Delta F/F$) of early time X-ray flares, as presented in Table \ref{tab:xrayfit}, are commonly observed in early X-ray light curves for both long and short duration GRBs \citep{chincarini2010, margutti2010, margutti2011, bernardini2011}.

\subsection{Optical light curve}
\label{sect:optana}

\par Figure \ref{fig:lc} shows the calibrated optical light curve (LC) of GRB~140430A, which is complex and could not be described by a simple power-law behavior. Overall, the light curve is qualitatively described by at least two long-lasting emission episodes joined by a plateau phase at $\sim 2000\,\mathrm{s}$. The excellent temporal sampling at early time, however, reveals additional components.

\par We fitted optical light curves with phenomenological models as typically used in the literature for optical afterglows. Possible theoretical models will be discussed in Section \ref{sect:discussion}. We used a combination of two Beuermann profiles (B), i.e., smoothly connected broken power-laws \citep{beuermann1999, guidorzi2014}, for broad components, and (following \citealt{kruehler2009}) two Gaussian profiles (G) corresponding to early time flares:
\begin{align}
\label{eq:beuermann}
F(t) &= \sum _{i=1}^2 B_i \times \frac{1+\alpha_\mathrm{R,i}/\alpha_\mathrm{D,i}}{(t/t_\mathrm{P,i})^{-\alpha_\mathrm{R,i}} + (\alpha_\mathrm{R,i}/\alpha_\mathrm{D,i})\,(t/t_\mathrm{P,i})^{\alpha _\mathrm{D,i}}} \notag \\
& + \sum _{j=1}^2 G_j \times \exp{\left(\frac{-(t - t_\mathrm{P,j})^2}{2 \sigma _j ^2} \right)},
\end{align}
where $B_i$ and $G_j$ are normalization constants, $t_\mathrm{P}$ are peak times, $\alpha _\mathrm{R}$ are rise indices and $\alpha _\mathrm{D}$ are decay indices. Parameters $\sigma _j$ from Gaussian profiles can be connected with the overall duration of the bump as $\Delta t \approx 2 \times \mathrm{FWHM} \approx 2 \times 2\sqrt{2 \ln 2} \, \sigma$. Note that we fixed the smoothing parameter from a more general Beuermann equation to $s = 1$, due to not very well constrained peak shapes of broad Beuermann components (using $s=0.5$ and $s=2$ does not change results). We also note that the Gaussian description of temporal profile is not physically motivated, but provides reasonable values of peak time and width.

\par We fitted all 3 wavelength bands simultaneously with different normalization factors for each color. For Beuermann profiles we assumed common $t_\mathrm{P, i}$, $\alpha _\mathrm{R,i}$, and $\alpha _\mathrm{D,i}$, while for Gaussian profiles we assumed different $t_\mathrm{P, j}$ and $\sigma _j$ for each color. We used Levenberg-Marquardt algorithm, assuming that photometric uncertainties are normally distributed. The fitting method provides the best-fit values and $1\sigma$ uncertainties of free parameters (Table \ref{tab:opticalfit}). The resulting $\chi ^2 = 168$ with with $\mathrm{d.o.f.} = 125$ has low P-value of $\mathrm{P} = 0.006$, indicating that the model does not describe well our complex dataset. Nevertheless, when testing different models by adding or removing Beuermann and/or Gaussian components, the fit did not improve and the residuals increased.

\begin{deluxetable}{lcccc}[!h]
\tablecaption{Optical LC best fit (2 Beuermann + 2 Gaussian) parameters}
\tabletypesize{\footnotesize}
\tablewidth{\linewidth}
\startdata
\hline \hline \\[-0.2cm]
\multicolumn{5}{c}{Beuermann peaks (underlying components):} \\
Peak & Peak time [s] & $\alpha _\mathrm{rise}$ & $\alpha _\mathrm{decay}$ & $F _\mathrm{p} \, [\mathrm{mJy}]$ \\
\hline \\[-0.2cm]
1, V & $\sim 260$ & $3.37 \pm 1.65$ & $0.97 \pm 0.12$ & $0.420$ \\
1, R & & & & $0.507$ \\
1, I & & & & $0.569$ \\ [0.2cm]
2, V & $\sim 2900$ & $2.73 \pm 0.44$ & $0.79 \pm 0.03$ & $0.095$ \\
2, R & & & & $0.115$ \\
2, I & & & & $0.129$ \\
\hline \\[0cm]
\multicolumn{5}{c}{Gaussian peaks (early time flares):} \\
Peak & Peak time [s] & Duration [s] & $\Delta t/t$ & $\Delta F/F$ \\
\hline \\[-0.2cm]
1, V & $168.4 \pm 2.7$ & $95.9 \pm 21.1$ & $0.57 \pm 0.13$ & $1.72$ \\
 & $\alpha _\mathrm{rise} = 6.2$ & $\alpha _\mathrm{decay} = 3.8$ & & \\
1, R & $166.5 \pm 4.5$ & $133.4 \pm 25.8$ & $0.80 \pm 0.16$ & $1.76$ \\
 & $\alpha _\mathrm{rise} = 5.0$ & $\alpha _\mathrm{decay} = 2.7$ & & \\
1, I & $170.6 \pm 6.7$ & $175.9 \pm 31.1$ & $1.03 \pm 0.19$ & $1.68$ \\ 
 & $\alpha _\mathrm{rise} = 4.2$ & $\alpha _\mathrm{decay} = 2.2$ & & \\[0.2 cm]
2, V & $239.7 \pm 3.8$ & $46.4 \pm 23.2$ & $0.19 \pm 0.10$ & $0.37$ \\
 & $\alpha _\mathrm{rise} = 4.4$ & $\alpha _\mathrm{decay} = 4.5$ & & \\
2, R & $257.8 \pm 5.5$ & $47.0 \pm 26.9$ & $0.18 \pm 0.10$ & $0.37$ \\
 & $\alpha _\mathrm{rise} = 4.2$ & $\alpha _\mathrm{decay} = 5.0$ & & \\
2, I & $259.3 \pm 7.5$ & $20.9 \pm 17.5$ & $0.08 \pm 0.07$ & $0.43$ \\
 & $\alpha _\mathrm{rise} = 8.7$ & $\alpha _\mathrm{decay} = 11.0$ & & 
\enddata
\tablecomments{The fit was performed simultaneously in V, R and I bands, with peak times, $\alpha _\mathrm{rise}$, $\alpha _\mathrm{decay}$ of the Beuermann profiles taken as common parameters among all 3 bands. Parameters for Gaussian profiles (peak time and width) were taken different for each band. $F_\mathrm{p}$ is the flux density at the time of the peak for each of two Beuermann profiles, and although the peak flux density is different for each band, we assumed a common flux density ratio between the first and the second Beuermann peak for all bands (the ratio obtained from the fit is $0.23 \pm 0.03$). $\Delta t/t$ is calculated as duration divided by peak time. $\Delta F/F$ is calculated as the ratio between difference of flux density at the time of the Gaussian peak and flux density of the underlying Beuermann peak, divided by flux density of the underlying Beuermann peak. Based on Figure \ref{fig:lc_derivative}, estimates of the maximum power-law rise ($\alpha _\mathrm{rise}$) and decay ($\alpha _\mathrm{decay}$) indices for Gaussian peaks are also given for each optical band.}
\label{tab:opticalfit}
\end{deluxetable} 

\par To obtain additional information from the fitted model, especially the power-law rise and decay indices, both data and best fit model are represented in differential plot in log-log scale (Figure \ref{fig:lc_derivative}). Although data show large scatter and relatively large uncertainties, which results in large scatter in differential plot, the best fit model shows that during the first optical flare the initial rising part power-law index is between $\alpha _\mathrm{rise} \sim 4$ and $\alpha _\mathrm{rise} \sim 6$ (depending on wavelength), followed by a decay with power-law index between $\alpha _\mathrm{decay} \sim 2$ and $\alpha _\mathrm{decay} \sim 4$. During the second optical flare, both rising and decaying indices are around $\sim 5$ (the I band fit is not very constraining for the second flare). The light curve then shows more smooth behavior with decaying power-law index $\alpha _\mathrm{decay} \sim 1$ from $\sim 300\,\mathrm{s}$ to $\sim 10^3\,\mathrm{s}$. We note that at that time, the afterglow has dimmed for more than $1.5 \,\mathrm{mag}$ and photometric uncertainties, especially for RINGO3 frames due to low sensitivity, become larger, but OSN I-band and STELLA r'-band points which are available at that time are of much better quality and thus more constraining. Differential plot between $\sim 1000\,\mathrm{s}$ and $\sim 3000\,\mathrm{s}$ thus shows a smooth transition from decay to a plateau phase, and then back to decay phase with a slightly shallower decay index $\alpha _\mathrm{decay} \sim 0.8$.

\begin{figure}[!ht]
\begin{center}
\includegraphics[width=1\linewidth]{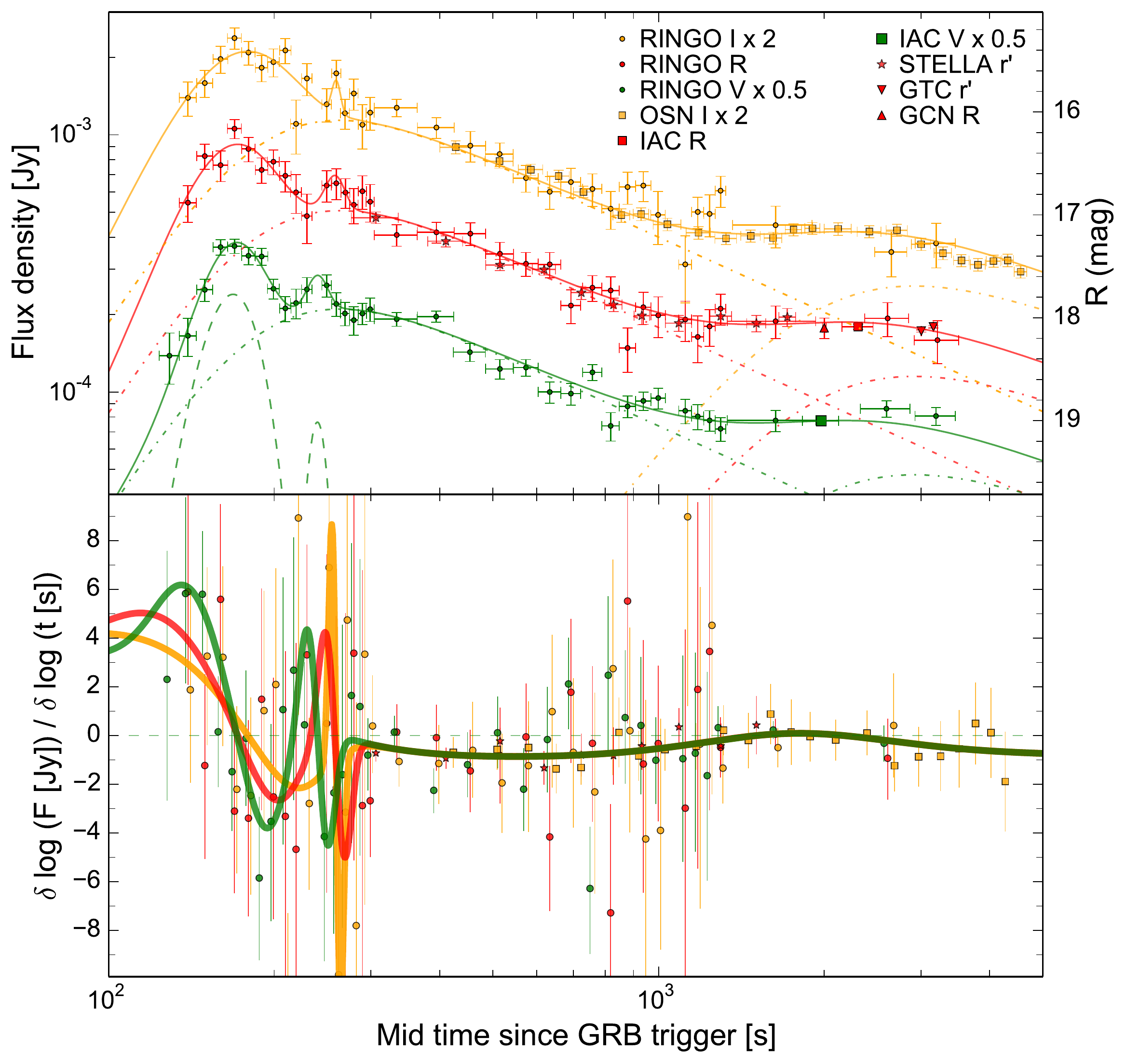} 
\caption{\label{fig:lc_derivative}GRB~140430A optical light curves in 3 bands, together with the derivative plot in the log-log scale. The derivative plot clearly shows deviations from a simple power-law behavior at early times, which corresponds to early time optical flares. For results obtained from this plot see Section \ref{sect:optana}.}
\end{center}
\end{figure}

\subsection{Broadband spectral energy distribution}
\label{sect:sedana}

\par At early times, due to simultaneous sampling of the light curve in 3 visible wavelength bands and in X-ray band, we could study the temporal evolution of spectral energy distribution (SED). The spectral index of the X-ray emission ($\beta_ \mathrm{X}$), obtained from the Swift Burst Analyser \citep{evans2010}, can be compared to the power-law slope of the optical SED ($\beta _ \mathrm{OPT}$), obtained from fitting a power-law to optical data points, or to the broadband optical to X-ray extrapolated spectral index ($\beta_ \mathrm{R-X}$), obtained from fitting a power-law to optical R-band and contemporaneous X-ray ($10\,\mathrm{keV}$) flux densities. We neglected the contribution of host galaxy extinction in optical bands, due to its low upper limit (see the end of this Section and Figure \ref{fig:sedlate}). SED evolution is presented in Figure \ref{fig:lc_beta}. 

\begin{figure}[!ht]
\begin{center}
\includegraphics[width=1\linewidth]{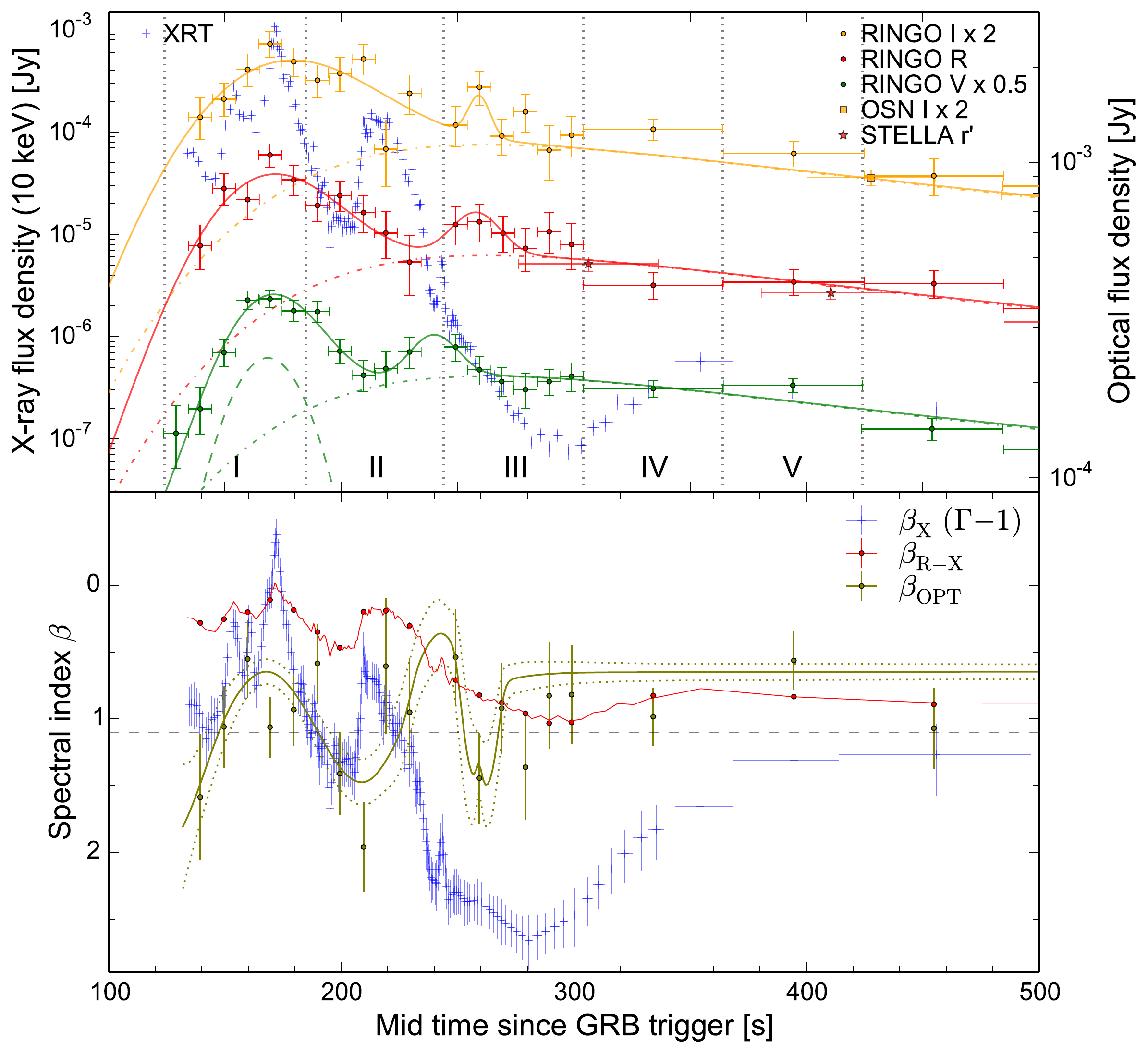} 
\caption{\label{fig:lc_beta}GRB~140430A early time X-ray and optical light curves in linear temporal scale (the optical flux density scale has been stretched and is represented on the right axis), together with the spectral power-law index of the X-ray emission ($\beta _\mathrm{X}$), optical emission ($\beta _\mathrm{OPT}$), and optical to X-ray extrapolated index ($\beta _\mathrm{R-X}$). Red solid line connecting $\beta _\mathrm{R-X}$ points is obtained by fitting a power-law SED to best fit R-band model and X-ray points. Olive solid line connecting $\beta _\mathrm{OPT}$ points is obtained by fitting a power-law SED to best fit optical models, and the corresponding dotted olive lines represent $1\sigma$ confidence interval of the fit. Dashed horizontal line corresponds to late time $\beta _\mathrm{X} \approx 1.1$. Vertical dotted lines in top panel show 5 time epochs where polarimetry has been performed (see Section \ref{sect:polarization} and Table \ref{tab:polarization}). For discussion on SED analysis see Section \ref{sect:sedana}.}
\end{center}
\end{figure}

\par We see from Figure \ref{fig:lc_beta} (bottom panel) that at early times the X-ray spectral index is highly variable and that it tracks the X-ray light curve throughout the flares. The hard-to-soft spectral evolution is observed. At the end of the last prominent X-ray flare at $\sim 240\,\mathrm{s}$ the X-ray emission becomes extremely soft ($\beta _\mathrm{X} \sim 2.5$) during the steep decay phase, but then it hardens at later times, when flaring is no longer present, and stays around the value of $\beta _\mathrm{X} = 1.10 \pm 0.04$, as obtained from the late time XRT spectral fit \citep{evans2009} and consistent with our late time SED fit (Figure \ref{fig:sedlate}). 

\par The temporal evolution of optical to X-ray extrapolated index ($\beta _\mathrm{R-X}$) and optical spectral index ($\beta _\mathrm{OPT}$) helps us understand if emission powering early time X-ray flares also manifests itself in the optical domain. The first thing that we notice is that at early times the variability of spectral index in broadband optical to X-ray SED is not so prominent as in the X-ray alone. The emission tends to be harder at the beginning, changing to softer at later times (after $\sim 400\,\mathrm{s}$), when it is consistent with $\beta _\mathrm{X}$. 

\par The temporal behavior of the optical spectral index ($\beta _\mathrm{OPT}$) suggests some degree of variability at early time but it is difficult to quantify this given the uncertainties. The statistical significance of variations in $\beta _\mathrm{OPT}$ is low, as the constant fit to $\beta _ \mathrm{OPT}$ over time interval $140-450\,\mathrm{s}$ gives an acceptable fit of average spectral index $\beta _\mathrm{OPT} = 0.97 \pm 0.08$, with reduced $\chi ^2$ of $1.4$ and P-value of $0.12$. After $300\,\mathrm{s}$ the spectral index implied by the best fitting model converges to $\beta _\mathrm{OPT} = 0.65 \pm 0.06$, while the variations implied by the RINGO3 data at later times become less constraining due to large uncertainties.

\par We built late time broadband SED at $\sim 3200\,\mathrm{s}$, when photometric data in 4 optical bands are available. We took X-ray data from XRT spectra repository \citep{evans2009} in time interval $3 - 30\,\mathrm{ks}$, in which there is no spectral evolution. The mean time was computed as $\sum _i (t_i \Delta t_i )/ \sum _i \Delta t_i $, where $t_i$ is the mid-time of individual exposure and $\Delta t_i$ is the exposure time. By knowing the temporal power-law slope, we normalized obtained flux densities to the epoch of optical observations. Using \texttt{XSPECv12.8}, we fitted the Milky Way, Large Magellanic Cloud, and Small Magellanic Cloud (SMC) average extinction curves \citep{pei1992} to our dataset, combined with a single or a broken power-law slope. The best fit was obtained using broken power-law and SMC extinction profile, which is most common for GRB SEDs (e.g., \citealt{japelj2015} and references therein). By fixing the difference of power-law slopes to $\Delta \beta = 0.5$ and fixing $N_\mathrm{H,X}^\mathrm{Gal.} = 2.13\times 10^{21} \, \mathrm{cm^{-2}}$ \citep{willingale2013}, we obtained the following best fit parameters: $\beta _1 = 0.68^{+0.13}_{-0.14}$, $\beta _2 = 1.18$, $\nu _\mathrm{break} = 160^{+800}_{-140} \times 10^{15} \,\mathrm{Hz}$, $N_\mathrm{H,X} < 8 \times 10^{21}\,\mathrm{cm ^{-2}}$ and $A_\mathrm{V} < 0.14$, with $\chi ^2 / \mathrm{d.o.f} = 24.8/21$. The resulting late time broadband SED is presented in Figure \ref{fig:sedlate}.

\begin{figure}[!h]
\begin{center}
\includegraphics[width=1\linewidth]{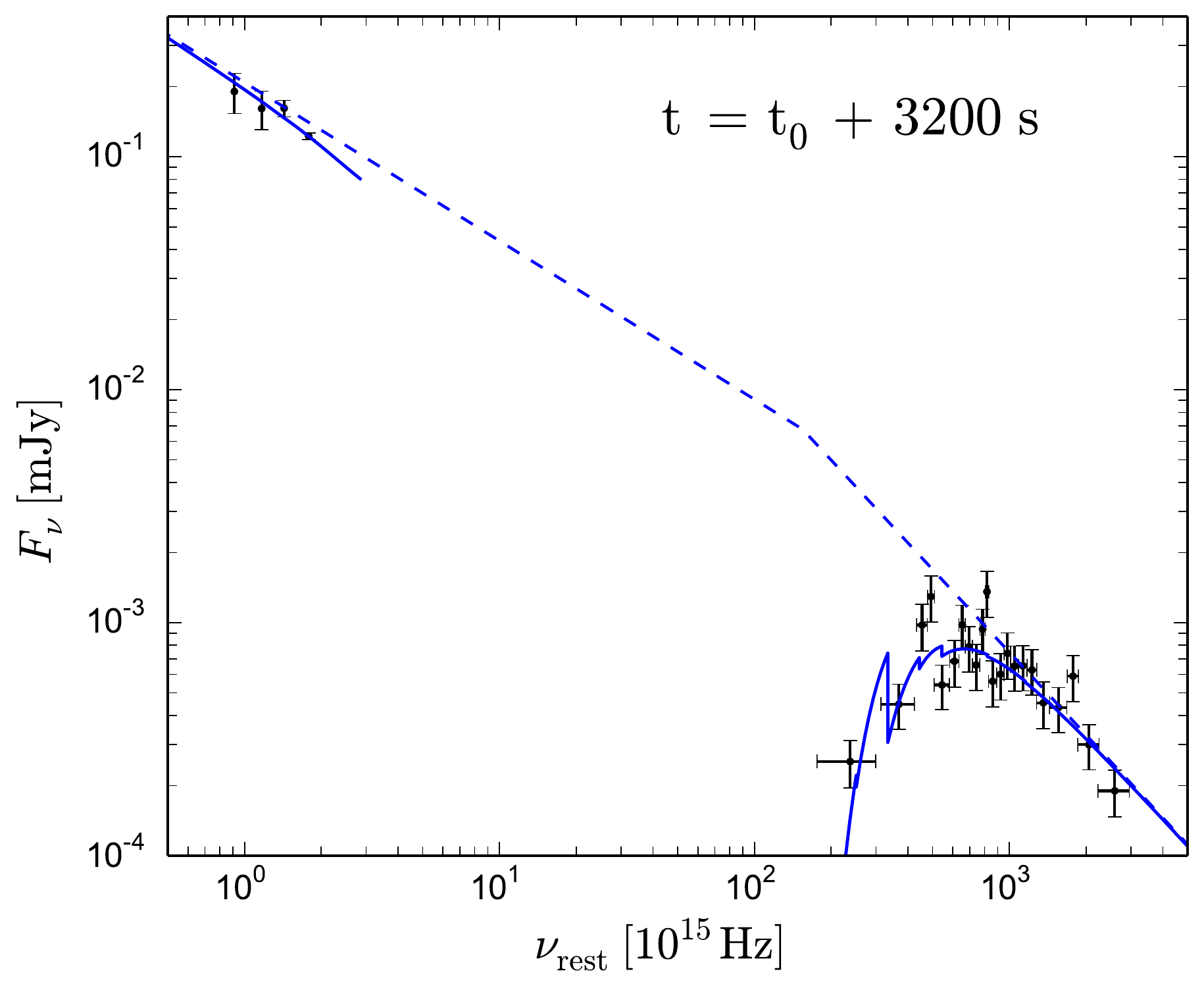} 
\caption{\label{fig:sedlate} GRB~140430A late time spectral energy distribution, with best fit results (see the text) plotted with blue lines. Solid line takes into account also optical extinction and soft X-ray absorption.}
\end{center}
\end{figure}

\subsection{Polarization}
\label{sect:polarization}

\par At the time of our measurements (30th April 2014), the RINGO3 polarimeter was in the process of commissioning. The data were taken when the full sensitivity of the instrument was not reached, resulting in low signal to noise (S/N) obtained from our measurement. Consequently, only the upper limits could be obtained for polarization degree $P$.

\par The polarimetry was done using standard RINGO procedures (Arnold et al., in prep.), similarly as for GRB~120308A \citep{mundell2013}, by first correcting the obtained Stokes parameters for instrumental-induced polarization and then by correcting the obtained polarization degree for instrumental depolarization. Corrections are obtained from the analysis of a full set of polarimetric standard stars. 

\par To correctly obtain the uncertainties ($\pm 1\sigma$) on the degree of polarization $P$, we performed a Monte Carlo simulation taking into account a normal distributed photometric uncertainties in each of 8 polaroid orientations. By generating $100,000$ simulated flux values, we calculated the corresponding normalized Stokes parameters `q' and `u' following \citet{clarke2002}. By fitting a 2D normal distribution to `qu' plane, we obtained the mean `q' and `u' values and $\sigma$ contours, which were then used to determine the upper limits on $P$. Figure \ref{fig:qu_plots} shows both the distribution of obtained $P$ values from this simulations and the `qu' scatter plots, for the RINGO3 epoch from $124$ - $244\,\mathrm{s}$. The final values on polarization degree for GRB~140430A at various RIGNO3 epochs from $124$ - $424\,\mathrm{s}$ are summarized in Table \ref{tab:polarization}.

\begin{figure*}[!ht]
\begin{center}
\includegraphics[width=0.33\linewidth]{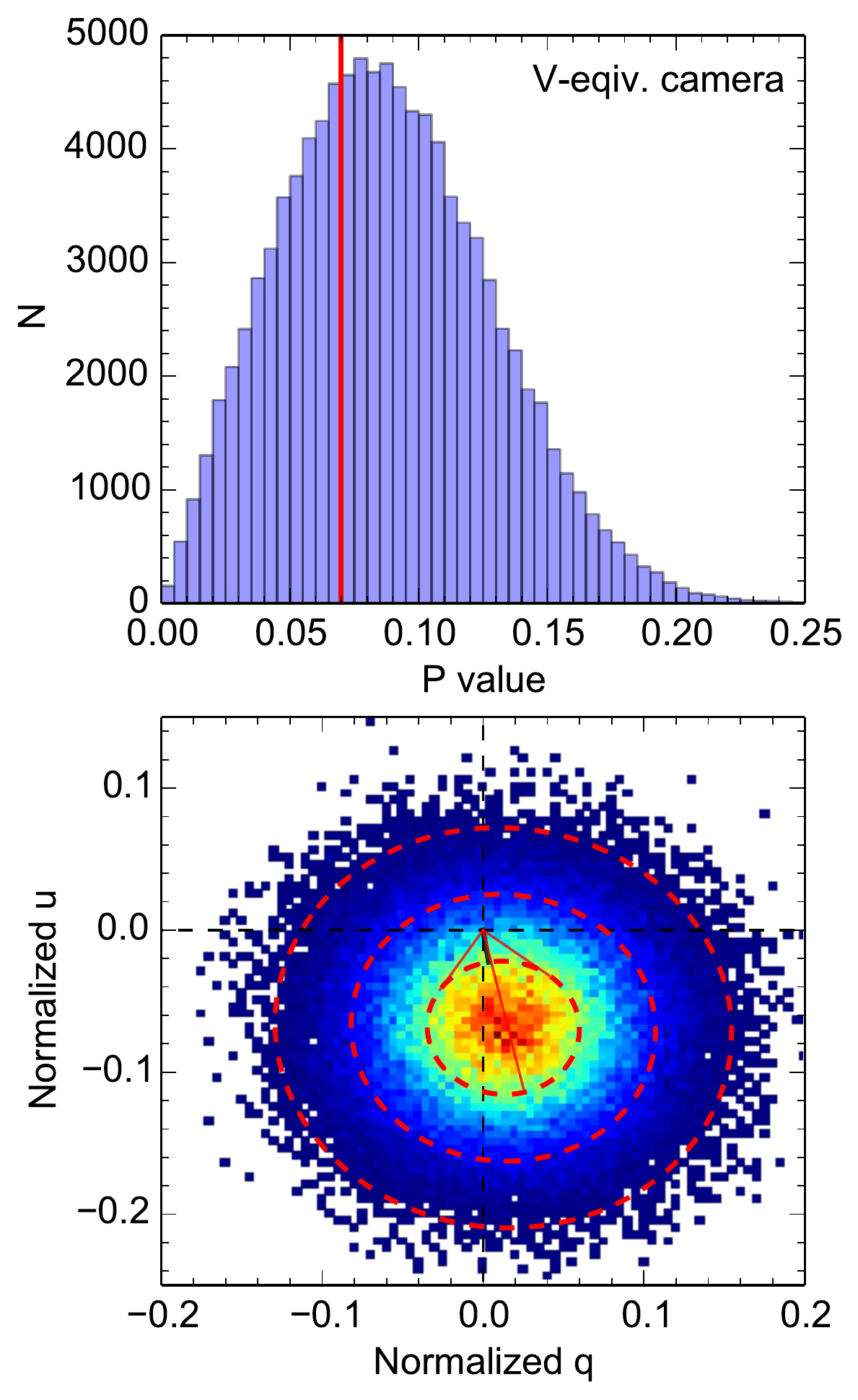} 
\includegraphics[width=0.33\linewidth]{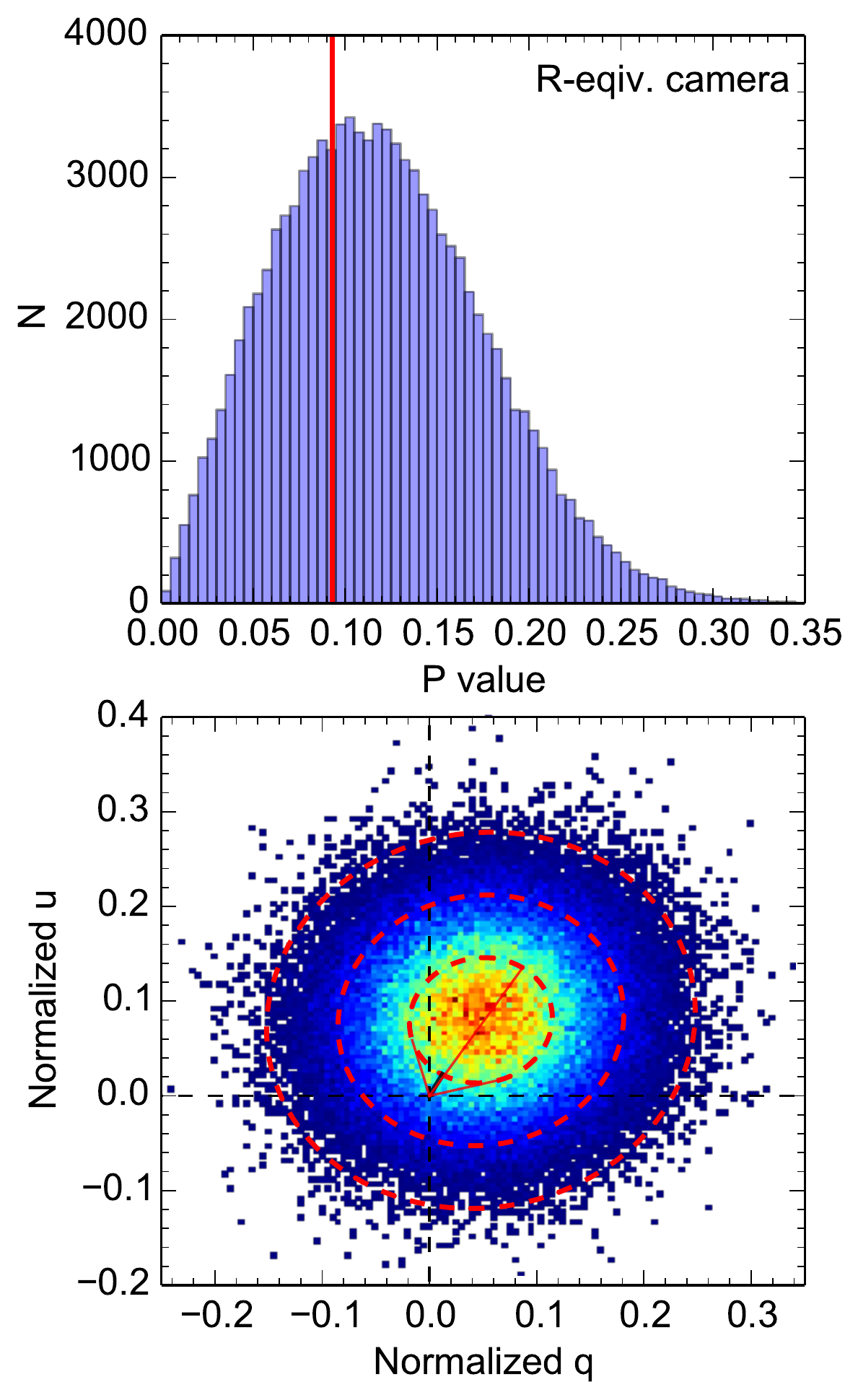} 
\includegraphics[width=0.33\linewidth]{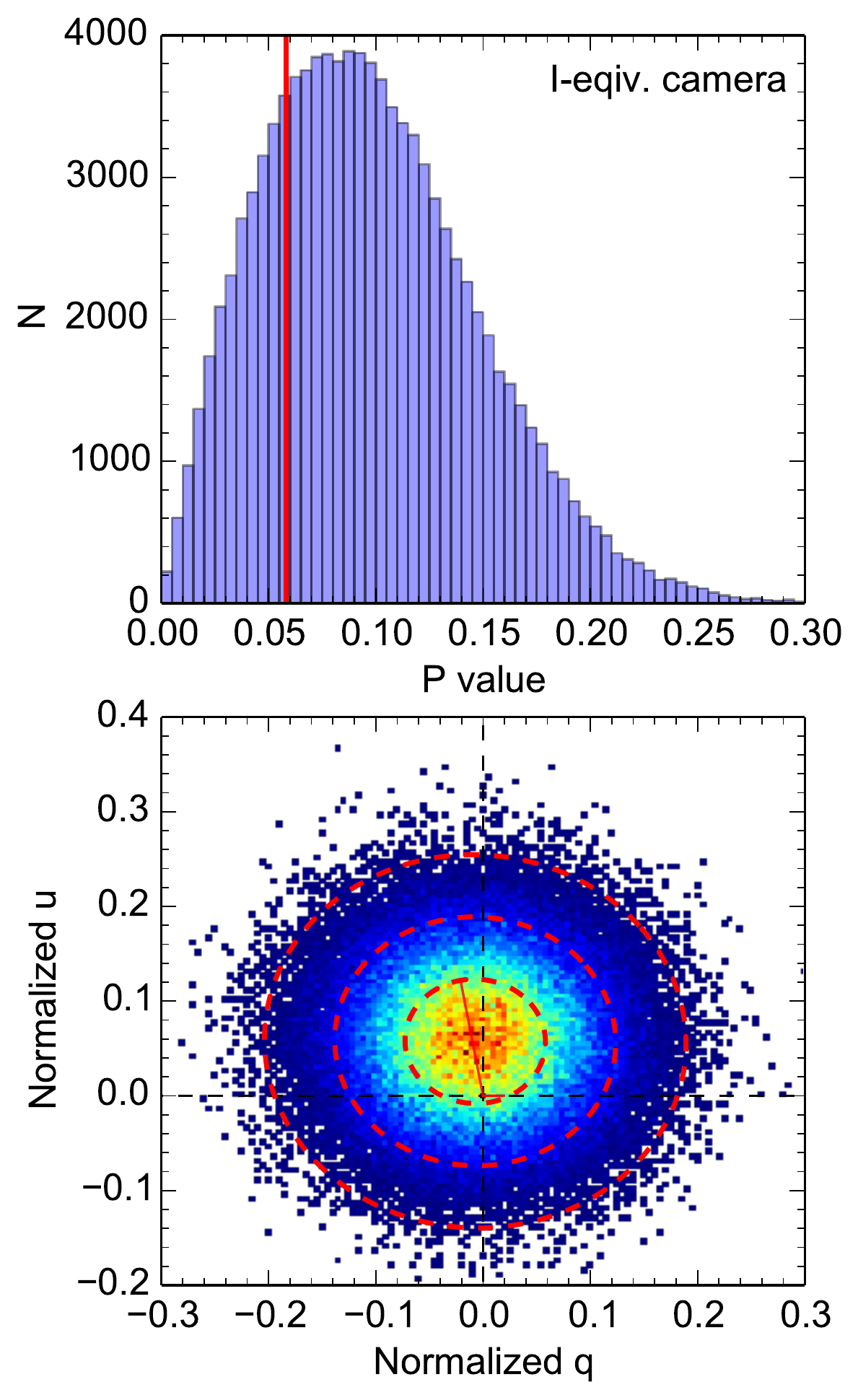} 
\caption{\label{fig:qu_plots} Results of Monte Carlo simulation to obtain uncertainties and upper limits on the degree of polarization $P$ for a $124$ - $244\,\mathrm{s}$ RINGO3 epoch. From left to right are results for 3 different cameras on RINGO3. On top are distributions of obtained $P$ values, with red solid lines indicating the measured $P$ when not taking flux uncertainties into account. On bottom are the scatter plots for normalized Stokes parameters `q' and `u', and red dashed lines indicate the 1$\sigma$, 2$\sigma$, and 3$\sigma$ contours (from center outwards), obtained by fitting a 2D normal distribution.}
\end{center}
\end{figure*}

\begin{deluxetable}{cccc}[!h]
\tablecaption{GRB~140430A optical polarization results}
\tabletypesize{\footnotesize}
\tablewidth{\linewidth}
\tablehead{
\colhead{Interval [s]} &
\colhead{V-eq. band} & 
\colhead{R-eq. band} &
\colhead{I-eq. band} 
}
\startdata
$124$ - $185$ & $< 19 \%$ & $< 19\%$ & $< 14\%$ \\
$185$ - $244$ & $< 20 \%$ & $< 16\%$ & $< 16\%$  \\
\vspace{-0.15cm}$244$ - $304$ & $< 22 \%$ & & $< 22 \%$  \\ \vspace{-0.15cm} 
 & & $< 12\%^\ast$ & \\
$304$ - $364$ & $< 23 \%$ & & $< 17 \%$  \\
$364$ - $424$ & $< 17 \%$ & $< 20\%$ & $< 10\%$ \\
& & & \\
$124$ - $244^\ast$ & $< 12\%$ & $< 16\%$ & $<12\%$ \\
 & $3\sigma <22\%$ & $3\sigma < 30\%$ & $3\sigma < 26\%$ 
\enddata
\tablecomments{$1\sigma$ (unless stated otherwise) upper limits on early time optical polarization degree $P$ from 3 RINGO3 cameras, in various time intervals. $^\ast$ indicates that data have been coadded in 2 time intervals, to obtain better S/N ratio.}
\label{tab:polarization}
\end{deluxetable}

\section{Discussion}
\label{sect:discussion}

\subsection{Early flares}
\label{sect:earlyflares}

\par We discuss the behavior of the light curves at early time in more detail and show that the initial optical flares appear more consistent with a prompt rather than afterglow origin.

\subsubsection{Prompt origin}
\label{sect:earlyflaresprompt}

\par The early optical light curve is dominated by highly variable components which are not easily explained in the pure context of standard external-shock afterglow scenario (either from forward- or reverse-shock), where smooth behavior is expected with rise index $\alpha_\mathrm{rise} < 5$ and decay index $\alpha_\mathrm{decay} < 2-3$ (e.g., \citealt{kobayashi2000, kobayashi2003, zhang2003}). Although temporal rise and decay indices for the first optical flare ($\alpha_\mathrm{rise} \sim 4.2 - 6.2$ and $\alpha_\mathrm{decay} \sim 2.2 - 3.8$, depending on the wavelength band) could be marginally consistent with those predicted by afterglow models, the indices are much steeper for the second optical flare ($\alpha_\mathrm{rise} \sim 4.2 - 8.7$ and $\alpha_\mathrm{decay} \sim 4.5 - 11.8$, see Section \ref{sect:optana}). The underlying broad Beuermann component point towards origin of the emission from external shock afterglow, but the superimposed optical components, which appear during the on-going high-energy gamma-ray and X-ray flares}, point at least partially towards internal shock origin. 

\par It has been suggested by studying X-ray flares that those can originate due to dissipation within internal shock region (e.g., \citealt{burrows2005, zhang2006, chincarini2007, troja2014}), similarly as gamma-ray emission \citep{reesmeszaros1992}. Recent statistical study of the waiting time distribution between gamma-ray pulses and X-ray flares showed that both phenomena are linked and likely produced by the same mechanism \citep{guidorzi2015}. Especially short variability timescales ($\Delta t/t$) and large amplitude variability ($\Delta F/F$) disfavor the origin of X-ray flares from the afterglow region due to density fluctuations, refreshed shocks or patchy shells \citep{ioka2005}. Values of $\Delta t/t$ and $\Delta F/F$ for early X-ray flares (Table \ref{tab:xrayfit}) fall outside of kinematically allowed regions for afterglow variability \citep{ioka2005}.

\par The study by \citet{kopac2013} showed that also optical emission at early times, especially when showing sharp and steep peaks in light curves, can originate from dissipation within internal shocks. Based on a simple two-shell internal shock collision model, the distribution of flux ratio between high-energy and optical emission can span from $(\nu F_\nu)^\gamma / (\nu F_\nu)^\mathrm{OPT} \gtrsim 1$ to $(\nu F_\nu)^\gamma / (\nu F_\nu)^\mathrm{OPT} \lesssim 10^5$, depending on various parameters like the bulk Lorentz factor of the ejected shell, energy density of electrons, energy density of magnetic fields, etc. Temporal delay of peaks at different energies can be due to different radii of shell collisions, depending on the initial separation between shells and the distribution of Lorentz factors. The flux ratio between X-ray and R-band for GRB~140430A is $(\nu F_\nu)^\mathrm{X} / (\nu F_\nu)^\mathrm{R} = 1.5 \times 10^4$ for the first optical flare, and $(\nu F_\nu)^\mathrm{X} / (\nu F_\nu)^\mathrm{R} = 2 \times 10^3$ for the second optical flare, indicating that the amount of energy emitted in optical bands is relatively small. Such flux ratio values are consistent with the values from the sample study of \citet{kopac2013}, and are comparable to for example GRB~080928 \citep{rossi2011} and GRB~110205A \citep{gendre2012, zheng2012}. Similarly as for other GRBs which show prompt optical flares, values of $\Delta t/t$ (Table \ref{tab:opticalfit}) are below $1$, and even below $0.2$ for the second optical flare.

\par Strong spectral evolution is commonly observed in time-resolved spectra of prompt gamma-ray emission (e.g., \citealt{lu2012}) and X-ray flares (e.g., \citealt{butlerkocevski2007, zhang2007}). Variability timescale of spectral behavior is short, similar to the corresponding light curve behavior, and different to that typically observed in the afterglow regime, where variations are usually smooth and the spectral slope stays constant or changes at breaks according to standard afterglow theory \citep{sari1998}. X-ray spectral index ($\beta _\mathrm{X}$) shows high variability during the prompt phase, and hard-to-soft spectral evolution which tracks the flares (see Figure \ref{fig:lc_beta}), pointing towards an internal shock origin. Variability in the optical spectral index ($\beta _\mathrm{OPT}$) at early time is also suggested by the data but large uncertainties (Figure \ref{fig:lc_beta}, olive points) prevents confirmation at a statistically significant level (see Section \ref{sect:sedana}). In contrast, the spectral index of the broadband optical to X-ray SED ($\beta _\mathrm{R-X}$) changes much more smoothly, with a gradual softening of the emission with time. This is likely due to the fact that early flares, which are much more powerful in the X-ray part of the spectrum, can mask the underlying synchrotron component from the afterglow emission, which is more prominent in the optical regime. 

\par Based on the discussion of a strong optical flare from GRB~080129 \citep{greiner2009}, likely causes for optical flares at early times could also be residual collisions \citep{li2008}, which predict variability on time scale of the same order as delay between gamma-ray and optical emission, or Poynting flux dissipation \citep{lyutikov2006,giannios2006}, which, in the case of GRB~140430A, is unlikely due to the lack of very high polarization.

\subsubsection{Afterglow origin}
\label{sect:earlyflaresRS}

\par
\begin{figure}[!ht]
\begin{center}
\includegraphics[width=1\linewidth]{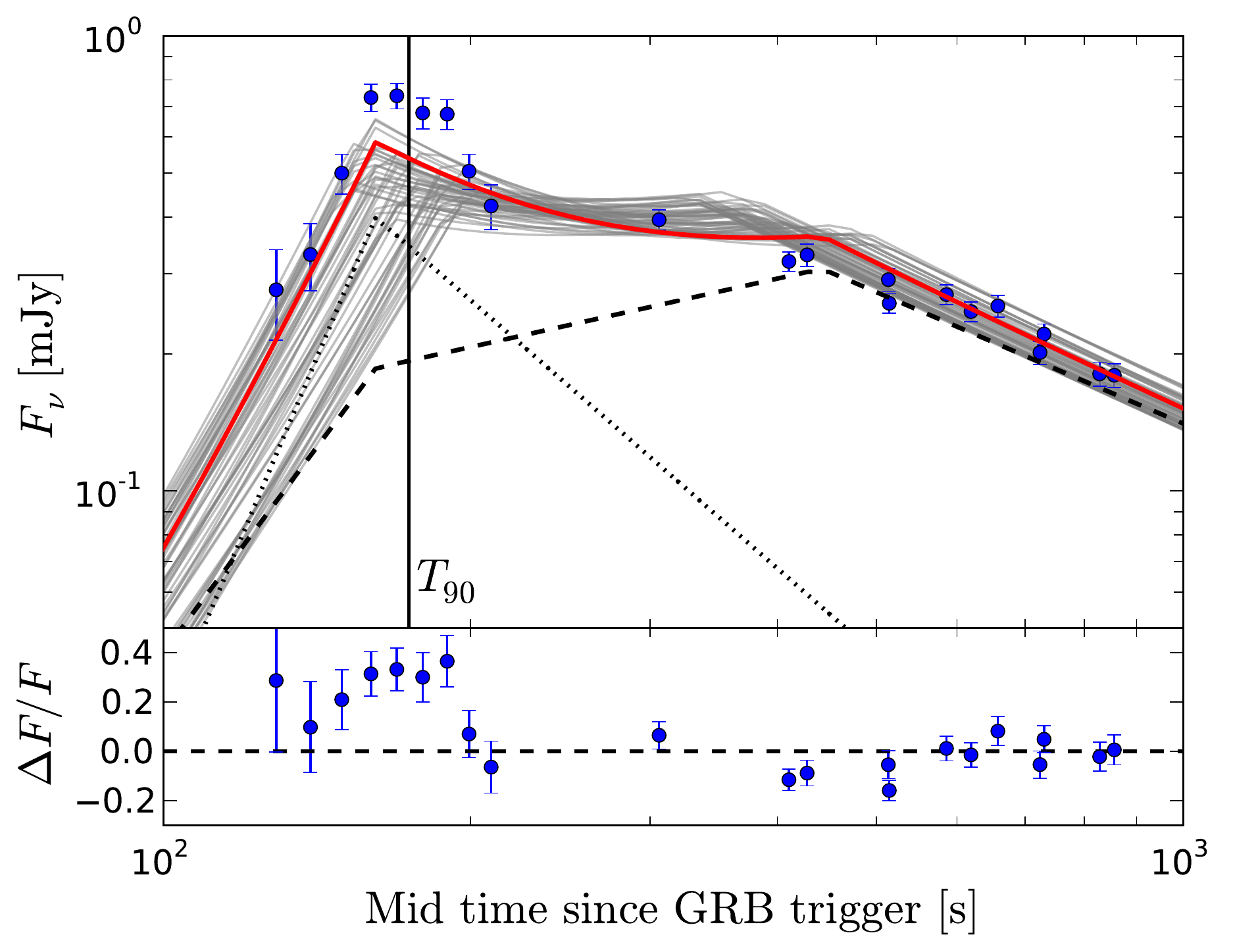} 
\caption{\label{fig:rsfslc}Top: combination of reverse-shock (dotted line) and forward-shock (dashed line) modeling for thin-shell case and constant ISM, on a modified optical light curve (see Section \ref{sect:earlyflaresRS}). Grey solid lines represent simulated models (see \citealt{japelj2014}), while best model is depicted with red solid line. Solid vertical line indicates $T_\mathrm{90}$. Bottom: residuals between data points and best model.}
\end{center}
\end{figure}

 An alternative scenario for the first optical flare is emission from an external shock, possibly reverse-shock emission. Examining the temporal behavior and following \citet{japelj2014}, we neglected the second optical flare, normalized the optical light curves to a common band using normalization parameters from Table \ref{tab:opticalfit}, and modeled the resultant dataset with a set of reverse- plus forward-shock light curves, assuming thin- or thick-shell limit and interstellar medium (ISM) environment of constant density \citep{kobayashi2000, zhang2003, japelj2014} (Figure \ref{fig:rsfslc}). The values emerging from the best model are $t_\mathrm{peak,FS} \approx 444\,\mathrm{s}$, $t_\mathrm{peak,RS} \approx 162\,\mathrm{s}$, $R_\mathrm{B} \equiv \epsilon _\mathrm{B,r} / \epsilon _\mathrm{B,f} \approx 2.5$, $p \approx 2.3$, $\epsilon _\mathrm{e} \approx 0.2$, $\Gamma _0 \sim 40$, however the modeling can not explain well the dataset, as indicated by the resultant residuals which suggest the presence of additional emission component during the first optical peak. Another contradiction comes from the fact that initial steep rise $\alpha \sim 5$ can only be explained by thin-shell case, but the fact that the duration of the burst ($\mathrm{T_{90}} \sim 174\,\mathrm{s}$) is larger than the peak time of the optical emission strongly suggests the thick-shell case \citep{zhang2003}. Furthermore, initial rise is too steep to be explained by the reverse-shock from the wind-type environment \citep{kobayashi2003}.

\par A complementary test of external shock emission is that of color evolution. In the context of reverse shock emission, no color evolution is expected because the reverse shock peak is typically attributed to the change in dynamics of the ejecta, rather than due to the passage of the spectral break, unless these two events coincide. However, combination of 2 peaks as presented in Figure \ref{fig:rsfslc} indicates that forward shock peak should be due to typical synchrotron passage, and change in spectral index from $\beta = -1/3$ to $\beta = (p-1)/2$ is expected \citep{sari1998}. Using the 3-band optical data for GRB~140430A, we searched for color evolution around the optical peak times. As can be seen in Figure \ref{fig:lc_beta}, no such color evolution as predicted by the theory is observed during the apparent forward shock peak.

\par Moreover, the origin of early optical peaks is not consistent with density fluctuations in the circumburst medium -- changes in temporal slopes during the optical flares would require prohibitively high density fluctuations to explain them \citep{nakargranot2007}. In summary, we therefore favor an internal shock origin for the early optical flares, as described in Section \ref{sect:earlyflaresprompt}

\subsection{Post-flaring afterglow}

\par After the initial flaring episode (after $300\,\mathrm{s}$), optical emission shows decay with power-law index $\alpha _\mathrm{decay} \approx 1$, which is consistent with the forward-shock afterglow origin. This is supported also by the late time broadband SED, which is best modeled using broken power-law (see Section \ref{sect:sedana}), with the difference of spectral indices $\Delta \beta \approx 0.5$, as expected for a cooling break \citep{sari1998}. Due to a cooling frequency lying between optical and X-ray, one would expect that X-ray light curve would decay steeper than optical, but this is not evident at late time, when X-ray and optical decay indices are comparable. The contradiction can be explained if the surrounding environment is not constant ISM medium, but stratified medium with circumburst medium density given by $n = A\,R^{-k}$ \citep{chevalierli2000}. In this case, temporal decay indices for forward shock emission are given by $\alpha _\mathrm{decay} = - (3p(4-k)+5k-12)/(4(4-k))$ and $\alpha _\mathrm{decay} = - (3p-2)/4$, before and after the cooling break, respectively. Similar temporal decay indices from different spectral regimes are thus possible in environments with $k \sim 4/3$.

\subsection{Late re-brightening: energy injection}
\label{sect:laterebright}

\par At $\sim 2000\,\mathrm{s}$ the optical light curve shows a transition from a power-law decline with $\alpha _\mathrm{decay} \sim 1$ to a plateau phase, followed by a power-law decline with $\alpha _\mathrm{decay} \sim 0.8$ (Table \ref{tab:opticalfit}). Consistent power-law decline after the plateau phase is also obtained from the X-ray light curve. This could be explained in the context of late time continued energy injection by a central engine (e.g., \citealt{reesmeszaros1998, zhangmeszaros2002, zhang2006}). Shallower decay index after the plateau phase implies gradual and continuous energy injection. Following \citet{sarimeszaros2000}, we can estimate that the change in optical decay slope of the forward shock corresponds to the change in the power-law exponent of the ejected mass distribution ($ M (> \gamma) \propto \gamma ^{-s}$) from $s \approx 1$ (instantaneous case) to $s \sim 2$, which is typical for moderate continuous energy injection.

\par Density bumps or voids in the surrounding medium can produce bumps in the afterglow light curves \citep{uhm2014}, but the decay index is expected to return to the same value as before the peak, which is not the case in this GRB (the difference is $\sim 0.2$). Similarly, multi-component jets, together with a possibility that an outflow is seen slightly off-axis, can produce light curves with additional re-brightening \citep{kumar2000}, but if the circumburst environment stays of the same type, the light curve behavior before and after the re-brightening should be similar. 

\subsection{Optical polarization upper limits} 

\par GRB polarimetry offers a direct probe of the structure of magnetic fields within the emission region. Late-time optical measurements taken hours to days after the burst show low values of linear polarization of a few percent, consistent with synchrotron emission from a tangled magnetic fields in the shocked interstellar medium (e.g., \citealt {lazzati2003, greiner2003, wiersema2012, wiersema2014}). Circular polarization was recently detected at the $0.6\%$ level, which although small, is larger than expected theoretically \citep{wiersema2014}. In contrast, measurements of the early afterglow taken hundreds of seconds after the burst, when the properties of the original fireball are still encoded in the emitted light, can show high linear polarizations up to $30\%$, particularly when a reverse shock can be identified in the light curve -- this is consistent with theoretical predictions for a large scale ordered magnetic flow advected from the central engine (\citealt{steele2009, uehara2012, mundell2013}). 

\par High degrees of gamma-ray polarization ($P \sim 40 - 60 \%$) have been measured during the prompt emission for a number of GRBs, using dedicated instruments on \textit{INTEGRAL} and \textit{IKAROS} satellites (e.g., \citealt{gotz2009, gotz2013, gotz2014, yonetoku2012}), despite the highly variable and peaked nature of prompt flares. The origin of prompt emission is still unknown but these polarization levels suggest magnetic fields play an important role. However, because gamma-ray polarization measurements are difficult to obtain, and often accompanied by large systematic uncertainties, the measurement of optical polarization at early times would provide additional constraints on the origin of the prompt emission.

\par GRB~140430A provides an unique opportunity, as its optical emission was detected during the on-going prompt emission. The upper limits we derive on linear optical polarization in the very early afterglow (see Section \ref{sect:polarization}) are relatively low at $P < 12-16 \%$ ($1\sigma$), or $P < 22-30 \%$ ($3\sigma$), during first optical flare. The polarization limit of $P \lesssim 20\%$ ($1\sigma$) during the second optical flare is less constrained due to S/N limitations. If early time optical flares originate from within the internal shock region in the jet and if the jet is threaded with large-scale and ordered magnetic fields, we would expect the emission to be highly polarized (up to $60\%$, depending on electron distribution). In contrast, the observed polarization could be reduced in a number of scenarios if (a) the magnetic field is tangled locally, (b) there exists a number of patches within a region of size $1/\Gamma$ where the magnetic field is ordered locally but our viewing angle at the time of observation is already larger due to smaller $\Gamma$ (e.g., \citealt{gruzinovwaxman1999, lyutikov2003, granotkonigl2003}), (c) there are multiple, unresolved but highly variable optical components that average temporally to a lower net value of polarization, or (d) emission from spatially distinct regions -- such as internal shock and the external reverse shock regions (see discussion in Section \ref{sect:earlyflares}) -- each with high polarization but different position angles are observed simultaneously, resulting in an apparent lower total polarization degree.

\par In summary, the total degree of early time optical polarization for GRB~140430A measured during the high-energy X-ray and gamma-ray flares is lower than the commonly obtained levels of prompt gamma-ray polarization measurements of other GRBs and lower than that attributed to reverse-shock emission in GRB~120308A \citep{mundell2013}. As described above, the intrinsic polarization of this GRB may be low or the vector averaging of different highly polarized components may explain the lower net observed polarization. Definitive interpretations of prompt gamma-ray emission will be possible when measurements of both prompt gamma-ray polarization and contemporaneous optical polarization are available for comparison in the same GRB.

\section{Conclusions}
\label{sect:conclusions}

\par In this paper, we present a detailed analysis of the early optical emission from GRB~140430A, which was observed with the fast polarimeter RINGO3 mounted on the Liverpool Telescope. Due to fast response of the instrumentation, we were able to obtain optical measurements in 3 color bands during the ongoing prompt gamma-ray emission and X-ray flaring episode. Our measurements and analysis show:
\begin{itemize}
\item The multi-color optical light curve at early time is complex, and best described by two optical flares superimposed on broad underlying afterglow component. The optical flares are temporally coincident with the high-energy X-ray and gamma-ray flares (prompt GRB emission), suggesting a central engine origin. Temporal and spectral analysis of the early time broadband dataset indicates emission from internal shock dissipation, dominating over external shock afterglow.
\item At $\sim 2000\,\mathrm{s}$, a late time re-brightening (plateau) occurs which is interpreted as due to late central engine activity (reactivation) with continuous energy injection, based on shallower post-plateau decay. Structured jet and density fluctuations in the circumburst medium are disfavored due to change in decay slope and due to no spectral variability.
\item Optical polarimetry in 3 bands during the on-going prompt emission showed that contemporaneous optical flares are not highly polarized, with the upper limits of as low as $P < 12 \%$ ($1\sigma$). Alternatively, time-averaging of multiple emission components (to obtain sufficient S/N for polarimetry) with different polarization behavior may be responsible for lowering the overall polarization estimate. 
\end{itemize} 

\par Time-resolved polarimetry during the prompt gamma-ray phase is vital to determine whether individual emission components in early optical light curves are polarized. Ultimately, observing the same GRBs simultaneously in optical and gamma-ray polarization will revolutionize our understanding of GRB emission mechanisms. 

\acknowledgements{
We thank the anonymous referee for valuable comments which improved the paper. DK acknowledges support from the Science and Technology Facilities Council (STFC). CGM acknowledges support from the Royal Society, the Wolfson Society and STFC. AJCT thanks the support of the Spanish Ministry Project AYA2012-39727-C03-01 and the excellent support from the OSN staff. The Liverpool Telescope is operated on the island of La Palma by Liverpool John Moores University in the Spanish Observatorio del Roque de los Muchachos of the Instituto de Astrofisica de Canarias with financial support from the UK Science and Technology Facilities Council. The Gran Telescopio Canarias (GTC) is installed in the Spanish Observatorio del Roque de los Muchachos of the Instituto de Astrofisica de Canarias, in the island of La Palma. This work made use of data obtained with the STELLA robotic telescopes in Tenerife, an AIP facility jointly operated by AIP and IAC. This work made use of data supplied by the UK Swift Science Data Centre at the University of Leicester. Swift mission is funded in the UK by STFC, in Italy by ASI, and in the USA by NASA.}

\LongTables
\renewcommand{\tabcolsep}{2pt}
\begin{deluxetable}{cccccc}
\tablecaption{GRB~140430A: Photometry.}
\tabletypesize{\scriptsize}
\tablehead{
\colhead{$t_\mathrm{mid}$ [s]} & 
\colhead{Exp [s]} &
\colhead{Telescope} & 
\colhead{Band} &
\colhead{Magnitude} &
\colhead{$F_\nu^\mathrm{OPT}$ [mJy]} 
}
\startdata
$139.4$ & $10$ & LT & \textit{I} & $16.63 \pm 0.17$ & $0.696 \pm 0.106$	\\
$149.6$ & $10$ & LT & \textit{I} & $16.48 \pm 0.13$ & $0.795 \pm 0.095$	\\
$159.8$ & $10$ & LT & \textit{I} & $16.25 \pm 0.13$ & $0.985 \pm 0.118$	\\
$169.4$ & $10$ & LT & \textit{I} & $16.04 \pm 0.1$ & $1.188 \pm 0.113$	\\
$179.6$ & $10$ & LT & \textit{I} & $16.18 \pm 0.11$ & $1.044 \pm 0.11$	\\
$189.8$ & $10$ & LT & \textit{I} & $16.33 \pm 0.13$ & $0.911 \pm 0.107$	\\
$199.4$ & $10$ & LT & \textit{I} & $16.28 \pm 0.14$ & $0.959 \pm 0.119$	\\
$209.6$ & $10$ & LT & \textit{I} & $16.16 \pm 0.12$ & $1.064 \pm 0.118$	\\
$219.2$ & $10$ & LT & \textit{I} & $16.9 \pm 0.26$ & $0.551 \pm 0.131$	\\
$229.4$ & $10$ & LT & \textit{I} & $16.44 \pm 0.16$ & $0.828 \pm 0.118$	\\
$249.2$ & $10$ & LT & \textit{I} & $16.69 \pm 0.16$ & $0.657 \pm 0.097$	\\
$259.4$ & $10$ & LT & \textit{I} & $16.39 \pm 0.13$ & $0.867 \pm 0.107$	\\
$269$ & $10$ & LT & \textit{I} & $16.77 \pm 0.15$ & $0.607 \pm 0.08$	\\
$279.2$ & $10$ & LT & \textit{I} & $16.58 \pm 0.15$ & $0.724 \pm 0.1$	\\
$289.4$ & $10$ & LT & \textit{I} & $16.9 \pm 0.22$ & $0.547 \pm 0.107$	\\
$299$ & $10$ & LT & \textit{I} & $16.77 \pm 0.16$ & $0.61 \pm 0.089$	\\
$334.1$ & $60$ & LT & \textit{I} & $16.71 \pm 0.08$ & $0.637 \pm 0.049$	\\
$394.5$ & $61$ & LT & \textit{I} & $16.91 \pm 0.1$ & $0.534 \pm 0.05$	\\
$454.7$ & $60$ & LT & \textit{I} & $17.09 \pm 0.15$ & $0.453 \pm 0.062$	\\
$514.1$ & $60$ & LT & \textit{I} & $17.17 \pm 0.11$ & $0.421 \pm 0.044$	\\
$574.1$ & $60$ & LT & \textit{I} & $17.4 \pm 0.13$ & $0.34 \pm 0.04$	\\
$634.1$ & $60$ & LT & \textit{I} & $17.54 \pm 0.16$ & $0.3 \pm 0.044$	\\
$692.9$ & $57$ & LT & \textit{I} & $17.44 \pm 0.14$ & $0.328 \pm 0.043$	\\
$758.1$ & $61$ & LT & \textit{I} & $17.51 \pm 0.16$ & $0.308 \pm 0.044$	\\
$818.3$ & $60$ & LT & \textit{I} & $17.71 \pm 0.18$ & $0.258 \pm 0.043$	\\
$878.3$ & $60$ & LT & \textit{I} & $17.49 \pm 0.16$ & $0.313 \pm 0.045$	\\
$938.3$ & $60$ & LT & \textit{I} & $17.48 \pm 0.15$ & $0.317 \pm 0.042$	\\
$998.1$ & $61$ & LT & \textit{I} & $17.78 \pm 0.26$ & $0.244 \pm 0.057$	\\
$1118.3$ & $60$ & LT & \textit{I} & $18.28 \pm 0.33$ & $0.157 \pm 0.046$	\\
$1178.3$ & $60$ & LT & \textit{I} & $17.74 \pm 0.15$ & $0.251 \pm 0.035$	\\
$1238.3$ & $60$ & LT & \textit{I} & $17.76 \pm 0.2$ & $0.246 \pm 0.046$	\\
$1296.5$ & $57$ & LT & \textit{I} & $17.53 \pm 0.15$ & $0.303 \pm 0.042$	\\
$1632.5$ & $537$ & LT & \textit{I} & $17.87 \pm 0.2$ & $0.223 \pm 0.041$	\\
$2650.1$ & $356$ & LT & \textit{I} & $18.14 \pm 0.22$ & $0.175 \pm 0.035$	\\
$3196.6$ & $537$ & LT & \textit{I} & $18.05 \pm 0.22$ & $0.189 \pm 0.037$	\\
$427.8$ & $55$ & OSN & \textit{I} & $17.1 \pm 0.06$ & $0.447 \pm 0.026$	\\
$513.8$ & $60$ & OSN & \textit{I} & $17.23 \pm 0.06$ & $0.395 \pm 0.023$	\\
$585.9$ & $60$ & OSN & \textit{I} & $17.31 \pm 0.05$ & $0.366 \pm 0.018$	\\
$658$ & $60$ & OSN & \textit{I} & $17.38 \pm 0.06$ & $0.346 \pm 0.019$		\\
$730.3$ & $60$ & OSN & \textit{I} & $17.53 \pm 0.06$ & $0.3 \pm 0.016$		\\
$855.9$ & $60$ & OSN & \textit{I} & $17.76 \pm 0.07$ & $0.243 \pm 0.015$ 	\\
$929.7$ & $60$ & OSN & \textit{I} & $17.75 \pm 0.06$ & $0.246 \pm 0.014$	\\
$1037.9$ & $120$ & OSN & \textit{I} & $17.85 \pm 0.06$ & $0.224 \pm 0.012$	\\
$1182.1$ & $120$ & OSN & \textit{I} & $17.93 \pm 0.06$ & $0.208 \pm 0.012$	\\
$1326.4$ & $120$ & OSN & \textit{I} & $17.98 \pm 0.06$ & $0.198 \pm 0.011$	\\
$1471.1$ & $120$ & OSN & \textit{I} & $17.96 \pm 0.06$ & $0.202 \pm 0.011$	\\
$1615.4$ & $120$ & OSN & \textit{I} & $17.98 \pm 0.06$ & $0.198 \pm 0.011$	\\
$1760.1$ & $120$ & OSN & \textit{I} & $17.9 \pm 0.06$ & $0.214 \pm 0.011$	\\
$1904.9$ & $120$ & OSN & \textit{I} & $17.89 \pm 0.06$ & $0.216 \pm 0.011$	\\
$2121.2$ & $240$ & OSN & \textit{I} & $17.89 \pm 0.06$ & $0.215 \pm 0.012$	\\
$2416.9$ & $240$ & OSN & \textit{I} & $17.92 \pm 0.06$ & $0.21 \pm 0.011$	\\
$2713.1$ & $240$ & OSN & \textit{I} & $17.9 \pm 0.06$ & $0.213 \pm 0.011$	\\
$3002.8$ & $240$ & OSN & \textit{I} & $18.04 \pm 0.06$ & $0.188 \pm 0.01$	\\
$3289.9$ & $240$ & OSN & \textit{I} & $18.13 \pm 0.06$ & $0.173 \pm 0.01$	\\
$3555$ & $230$ & OSN & \textit{I} & $18.2 \pm 0.06$ & $0.162 \pm 0.009$	\\
$3808.7$ & $240$ & OSN & \textit{I} & $18.24 \pm 0.06$ & $0.156 \pm 0.009$	\\
$4064.6$ & $240$ & OSN & \textit{I} & $18.2 \pm 0.06$ & $0.161 \pm 0.009$	\\
$4313.5$ & $210$ & OSN & \textit{I} & $18.2 \pm 0.06$ & $0.163 \pm 0.009$	\\
$4553.9$ & $240$ & OSN & \textit{I} & $18.31 \pm 0.06$ & $0.147 \pm 0.008$	\\
 & & & & & \\
$37.8$ & $16$ & BOOTES & \textit{r'} & $16.14 \pm 0.15$ & $1.711 \pm 0.235$ \\
$79.8$ & $17$ & BOOTES & \textit{r'} & $16.48 \pm 0.14$ & $1.25 \pm 0.16$ \\ 
$139.4$ & $10$ & LT & \textit{R} & $17.19 \pm 0.18$ & $0.545 \pm 0.088$	\\
$149.6$ & $10$ & LT & \textit{R} & $16.73 \pm 0.12$ & $0.827 \pm 0.093$	\\
$159.8$ & $10$ & LT & \textit{R} & $16.82 \pm 0.15$ & $0.762 \pm 0.104$	\\
$169.4$ & $10$ & LT & \textit{R} & $16.46 \pm 0.09$ & $1.056 \pm 0.09$	\\
$179.6$ & $10$ & LT & \textit{R} & $16.66 \pm 0.12$ & $0.881 \pm 0.096$	\\
$189.8$ & $10$ & LT & \textit{R} & $16.87 \pm 0.12$ & $0.73 \pm 0.082$	\\
$199.4$ & $10$ & LT & \textit{R} & $16.79 \pm 0.12$ & $0.786 \pm 0.087$	\\
$209.6$ & $10$ & LT & \textit{R} & $16.93 \pm 0.15$ & $0.693 \pm 0.094$	\\
$219.2$ & $10$ & LT & \textit{R} & $17.09 \pm 0.19$ & $0.597 \pm 0.102$	\\
$229.4$ & $10$ & LT & \textit{R} & $17.33 \pm 0.24$ & $0.483 \pm 0.105$	\\
$249.2$ & $10$ & LT & \textit{R} & $17.02 \pm 0.15$ & $0.635 \pm 0.088$	\\
$259.4$ & $10$ & LT & \textit{R} & $17 \pm 0.15$ & $0.648 \pm 0.089$	\\
$269.6$ & $10$ & LT & \textit{R} & $17.09 \pm 0.14$ & $0.596 \pm 0.079$	\\
$279.2$ & $10$ & LT & \textit{R} & $17.21 \pm 0.17$ & $0.534 \pm 0.083$	\\
$289.4$ & $10$ & LT & \textit{R} & $17.08 \pm 0.16$ & $0.603 \pm 0.09$	\\
$299$ & $10$ & LT & \textit{R} & $17.18 \pm 0.18$ & $0.549 \pm 0.092$	\\
$334.1$ & $60$ & LT & \textit{R} & $17.5 \pm 0.11$ & $0.408 \pm 0.039$	\\
$394.5$ & $61$ & LT & \textit{R} & $17.47 \pm 0.1$ & $0.418 \pm 0.038$	\\
$454.7$ & $60$ & LT & \textit{R} & $17.49 \pm 0.11$ & $0.413 \pm 0.042$	\\
$514.7$ & $60$ & LT & \textit{R} & $17.68 \pm 0.12$ & $0.345 \pm 0.037$	\\
$574.1$ & $60$ & LT & \textit{R} & $17.78 \pm 0.12$ & $0.315 \pm 0.033$	\\
$634.1$ & $60$ & LT & \textit{R} & $17.78 \pm 0.12$ & $0.314 \pm 0.035$	\\
$692.9$ & $57$ & LT & \textit{R} & $18.19 \pm 0.16$ & $0.217 \pm 0.033$	\\
$758.1$ & $61$ & LT & \textit{R} & $18.01 \pm 0.12$ & $0.254 \pm 0.029$	\\
$818.3$ & $60$ & LT & \textit{R} & $18.04 \pm 0.14$ & $0.248 \pm 0.032$	\\
$878.3$ & $60$ & LT & \textit{R} & $18.61 \pm 0.22$ & $0.148 \pm 0.029$	\\
$938.3$ & $60$ & LT & \textit{R} & $18.21 \pm 0.16$ & $0.214 \pm 0.03$	\\
$998.7$ & $61$ & LT & \textit{R} & $18.29 \pm 0.2$ & $0.199 \pm 0.036$	\\
$1118.3$ & $60$ & LT & \textit{R} & $18.33 \pm 0.2$ & $0.191 \pm 0.034$	\\
$1178.3$ & $60$ & LT & \textit{R} & $18.51 \pm 0.23$ & $0.164 \pm 0.034$	\\
$1238.3$ & $60$ & LT & \textit{R} & $18.4 \pm 0.18$ & $0.18 \pm 0.029$	\\
$1296.5$ & $57$ & LT & \textit{R} & $18.22 \pm 0.15$ & $0.211 \pm 0.029$	\\
$1632.4$ & $597$ & LT & \textit{R} & $18.34 \pm 0.16$ & $0.188 \pm 0.027$	\\
$2610$ & $478$ & LT & \textit{R} & $18.32 \pm 0.17$ & $0.193 \pm 0.029$	\\
$3215.1$ & $597$ & LT & \textit{R} & $18.53 \pm 0.21$ & $0.159 \pm 0.03$	\\
$306.2$ & $60$ & STELLA & \textit{r'} & $17.33 \pm 0.06$ & $0.476 \pm 0.025$	\\
$410.5$ & $60$ & STELLA & \textit{r'} & $17.55 \pm 0.05$ & $0.386 \pm 0.019$	\\
$514.9$ & $60$ & STELLA & \textit{r'} & $17.78 \pm 0.05$ & $0.312 \pm 0.016$	\\
$619.3$ & $60$ & STELLA & \textit{r'} & $17.83 \pm 0.05$ & $0.299 \pm 0.015$	\\
$723.7$ & $60$ & STELLA & \textit{r'} & $18.05 \pm 0.06$ & $0.243 \pm 0.014$	\\
$828.1$ & $60$ & STELLA & \textit{r'} & $18.17 \pm 0.07$ & $0.218 \pm 0.013$	\\
$932.3$ & $60$ & STELLA & \textit{r'} & $18.28 \pm 0.09$ & $0.198 \pm 0.016$	\\
$1088.8$ & $120$ & STELLA & \textit{r'} & $18.35 \pm 0.08$ & $0.185 \pm 0.013$	\\
$1297.5$ & $120$ & STELLA & \textit{r'} & $18.28 \pm 0.07$ & $0.197 \pm 0.012$	\\
$1506.3$ & $120$ & STELLA & \textit{r'} & $18.35 \pm 0.08$ & $0.185 \pm 0.014$	\\
$1714.9$ & $120$ & STELLA & \textit{r'} & $18.29 \pm 0.09$ & $0.195 \pm 0.016$	\\
$2001$ & $30$ & LT-IO:O$^\ast$ & \textit{R} & $18.4 \pm 0.1$ & $0.177 \pm 0.016$	\\
$5240$ & $100$ & NOT$^\ast$ & \textit{R} & $18.78 \pm 0.02$ & $0.125 \pm 0.002$	\\
$24516$ & $600$ & VATT$^\ast$ & \textit{R} & $20.07 \pm 0.03$ & $0.038 \pm 0.001$	\\
$110916$ & $1500$ & VATT$^\ast$ & \textit{R} & $21.7 \pm 0.1$ & $0.008 \pm 0.001$	\\
$3004.3$ & $10$ & GTC & \textit{r'} & $18.62 \pm 0.01$ & $0.173 \pm 0.002$	\\
$3160.9$ & $10$ & GTC & \textit{r'} & $18.62 \pm 0.01$ & $0.179 \pm 0.002$	\\
$2305.2$ & $300$ & IAC & \textit{R} & $18.38 \pm 0.03$ & $0.179 \pm 0.004$	\\
 & & & & & \\
$129$ & $11$ & LT & \textit{V} & $18.2 \pm 0.25$ & $0.277 \pm 0.062$	\\
$139.4$ & $10$ & LT & \textit{V} & $18 \pm 0.18$ & $0.331 \pm 0.055$	\\
$149.6$ & $10$ & LT & \textit{V} & $17.54 \pm 0.11$ & $0.499 \pm 0.05$	\\
$159.8$ & $10$ & LT & \textit{V} & $17.12 \pm 0.08$ & $0.732 \pm 0.05$	\\
$169.4$ & $10$ & LT & \textit{V} & $17.11 \pm 0.07$ & $0.739 \pm 0.048$	\\
$179.6$ & $10$ & LT & \textit{V} & $17.21 \pm 0.08$ & $0.677 \pm 0.052$	\\
$189.8$ & $10$ & LT & \textit{V} & $17.21 \pm 0.08$ & $0.673 \pm 0.051$	\\
$199.4$ & $10$ & LT & \textit{V} & $17.53 \pm 0.1$ & $0.504 \pm 0.045$	\\
$209.6$ & $10$ & LT & \textit{V} & $17.72 \pm 0.12$ & $0.423 \pm 0.048$	\\
$219.2$ & $10$ & LT & \textit{V} & $17.67 \pm 0.14$ & $0.444 \pm 0.059$	\\
$229.4$ & $10$ & LT & \textit{V} & $17.53 \pm 0.12$ & $0.501 \pm 0.056$	\\
$249.2$ & $10$ & LT & \textit{V} & $17.49 \pm 0.11$ & $0.52 \pm 0.052$	\\
$259.4$ & $10$ & LT & \textit{V} & $17.68 \pm 0.12$ & $0.44 \pm 0.046$	\\
$269$ & $10$ & LT & \textit{V} & $17.77 \pm 0.11$ & $0.404 \pm 0.042$	\\
$279.2$ & $10$ & LT & \textit{V} & $17.84 \pm 0.14$ & $0.381 \pm 0.048$	\\
$289.4$ & $10$ & LT & \textit{V} & $17.77 \pm 0.1$ & $0.404 \pm 0.038$	\\
$299$ & $10$ & LT & \textit{V} & $17.73 \pm 0.11$ & $0.42 \pm 0.044$	\\
$334.1$ & $60$ & LT & \textit{V} & $17.82 \pm 0.07$ & $0.384 \pm 0.023$	\\
$394.1$ & $60$ & LT & \textit{V} & $17.79 \pm 0.05$ & $0.393 \pm 0.019$	\\
$453.9$ & $61$ & LT & \textit{V} & $18.14 \pm 0.09$ & $0.286 \pm 0.023$	\\
$514.7$ & $60$ & LT & \textit{V} & $18.31 \pm 0.09$ & $0.246 \pm 0.021$	\\
$574.1$ & $60$ & LT & \textit{V} & $18.29 \pm 0.08$ & $0.249 \pm 0.018$	\\
$634.1$ & $60$ & LT & \textit{V} & $18.53 \pm 0.1$ & $0.2 \pm 0.019$	\\
$692.9$ & $57$ & LT & \textit{V} & $18.55 \pm 0.11$ & $0.197 \pm 0.019$	\\
$758.1$ & $61$ & LT & \textit{V} & $18.34 \pm 0.08$ & $0.239 \pm 0.017$	\\
$818.3$ & $60$ & LT & \textit{V} & $18.86 \pm 0.14$ & $0.148 \pm 0.019$	\\
$878.3$ & $60$ & LT & \textit{V} & $18.67 \pm 0.11$ & $0.176 \pm 0.017$	\\
$938.3$ & $60$ & LT & \textit{V} & $18.62 \pm 0.09$ & $0.185 \pm 0.015$	\\
$998.7$ & $61$ & LT & \textit{V} & $18.59 \pm 0.1$ & $0.19 \pm 0.018$	\\
$1118.3$ & $60$ & LT & \textit{V} & $18.72 \pm 0.12$ & $0.169 \pm 0.018$	\\
$1178.3$ & $60$ & LT & \textit{V} & $18.77 \pm 0.12$ & $0.161 \pm 0.018$	\\
$1238.3$ & $60$ & LT & \textit{V} & $18.81 \pm 0.1$ & $0.155 \pm 0.014$	\\
$1296.5$ & $57$ & LT & \textit{V} & $18.89 \pm 0.12$ & $0.144 \pm 0.015$	\\
$1632.4$ & $597$ & LT & \textit{V} & $18.81 \pm 0.1$ & $0.155 \pm 0.015$	\\
$2596.7$ & $537$ & LT & \textit{V} & $18.69 \pm 0.08$ & $0.172 \pm 0.013$	\\
$3195.4$ & $537$ & LT & \textit{V} & $18.76 \pm 0.09$ & $0.161 \pm 0.013$	\\
$1826.4$ & $300$ & IAC & \textit{V} & $18.81 \pm 0.03$ & $0.155 \pm 0.004$	\\
 & & & & & \\
$1647.6$ & $300$ & IAC & \textit{B} & $19.24 \pm 0.03$ & $0.135 \pm 0.004$	
\enddata
\tablecomments{\label{tab:phot140430A}Magnitudes are not corrected for the Galactic extinction, while flux densities are. $^\ast$ indicates data obtained from GCN Circulars.}
\end{deluxetable}

\end{document}